\documentclass[12pt]{article}

\usepackage{epsf,epsfig}
\usepackage{color,colortbl}
\usepackage{graphics}
\usepackage{axodraw}

\renewcommand{\arraystretch}{1.2}

\def\slash#1{#1 \hskip-0.45em /}
\def\Slash#1{#1 \hskip-0.59em /}
\def\simgt{\rlap{\lower 3.5 pt \hbox{$\mathchar \sim$}} \raise 1pt \hbox {$>$}}
\def\simlt{\rlap{\lower 3.5 pt \hbox{$\mathchar \sim$}} \raise 1pt \hbox {$<$}}

\def\be{\begin{equation}}
\def\ee{\end{equation}}
\def\beq{\begin{eqnarray}}
\def\eeq{\end{eqnarray}}
\newcommand{\bea}{\begin{eqnarray}}
\newcommand{\eea}{\end{eqnarray}}
\newcommand{\beas}{\begin{eqnarray*}}
\newcommand{\eeas}{\end{eqnarray*}}

\newcommand{\ub}{\bar u}
\newcommand{\vb}{\bar v}

\newcommand{\cH}{{\cal H}}

\newcommand{\npslash}{{n \!\!\! /}_+ }
\newcommand{\nmslash}{{n \!\!\! /}_- }
\newcommand{\nph}{{\frac{\npslash}{2}}}
\newcommand{\nmh}{{\frac{\nmslash}{2}}}

\newcommand{\xib}{{\bar \xi}}

\newcommand{\hc}{{\mathrm{hc}}}

\newcommand{\Wcp}{W_{c2}}

\newcommand{\cO}{{\cal O}}

\newcommand{\sceti}{{SCET$_{\mathrm{I}}$}}
\newcommand{\scetii}{{SCET$_{\mathrm{II}}$}}

\newcommand{\eff}{{\mathrm{eff}}}

\newcommand{\eq}[1]{(\ref{#1})}

\begin{document}

\begin{titlepage}

\begin{flushright}
{\small
PITHA~06/09\\
hep-ph/0610322
}
\end{flushright}

\vspace{1.3cm}
\begin{center}
\Large\bf\boldmath
Spectator scattering at NLO in non-leptonic\\ $B$ decays: 
Leading penguin amplitudes
\unboldmath
\end{center}

\vspace{0.8cm}
\begin{center}
{\sc M.~Beneke and S.~J\"ager{\begingroup\def\thefootnote{\fnsymbol{footnote}}\footnote[1]{Address after 01 October 2006:
Arnold Sommerfeld Center, Department f\"ur Physik,
Ludwig-Maximilians-Universit\"at M\"unchen, Theresienstra{\ss}e 37,
D-80333 M\"unchen, Germany}\endgroup} } \\
\vspace{0.7cm}
{\sl Institut f\"ur Theoretische Physik E, RWTH Aachen\\
D--52056 Aachen, Germany}
\end{center}

\vspace{1.3cm}
\begin{abstract}
\vspace{0.2cm}\noindent
We complete the computation of the 1-loop ($\alpha_s^2$)
corrections to hard spectator scattering in non-leptonic $B$ decays
at leading power in $\Lambda/m_b$ by evaluating the penguin
amplitudes. This extends the knowledge of these
next-to-next-to-leading-order contributions
in the QCD factorization formula for $B$ decays to 
a much wider class of final states, including all pseudoscalar-pseudoscalar,
pseudoscalar-vector, and longitudinally polarized vector-vector 
final states, except final states with $\eta$ or $\eta^\prime$ mesons. 
The new 1-loop correction is significant for the colour-suppressed 
amplitudes, but turns out to be strongly suppressed for the 
leading QCD penguin amplitude $\alpha_4^p$. We provide numerical 
values of the phenomenological $P/T$ and $C/T$ amplitude ratios 
for the $\pi\pi$, $\pi\rho$ and $\rho\rho$ final states, and discuss 
corrections to several relations between electroweak penguin and tree 
amplitudes.
\end{abstract}

\vfil
\end{titlepage}

\section{Introduction}

In the QCD factorization framework~\cite{Beneke:1999br,Beneke:2001ev} 
the matrix elements of the effective weak interaction operators relevant to 
charmless non-leptonic $B$ decays $B\to M_1 M_2$ take 
the (schematic) expression
\be                    
  \langle M_1 M_2 | Q_i | B \rangle
        = F^{B M_1}(0)\,\, T^\mathrm{I}_{i} * f_{M_2} \phi_{M_2}
          + T^\mathrm{II}_{i} * f_B \phi_{B+} * f_{M_1} \phi_{M_1} *
          f_{M_2} \phi_{M_2}
\label{eq:qcdfact}
\ee
at leading order in the $1/m_b$ expansion,
where $F^{B M_1}(0)$ denotes a $B\to M_1$ form factor at $q^2=0$, 
$f_M$ decay constants, and $\phi_M$ light-cone distribution
amplitudes. The convolution kernels $T_i^{\rm I,II}$ are 
short-distance, and can be expanded in a perturbation series in the 
strong coupling $\alpha_s$. Precise calculations of these kernels 
are required to make the framework predictive. This is particularly 
the case for direct CP asymmetries, since at leading order 
in the $1/m_b$ expansion strong interaction 
phases arise only from these kernels, and only through loop 
diagrams. 

The kernels $T_i^{\rm I}$ are currently known  at order 
$\alpha_s$~\cite{Beneke:1999br,Beneke:2001ev}, 
which includes a 1-loop correction to 
``naive factorization''. The situation is different for $T_i^{\rm II}$,
which always involves the exchange of a hard-collinear gluon with virtuality 
$m_b \Lambda$ ($\Lambda$ the strong interaction scale) with the
spectator quark in the $B$ meson.
Due to the presence of this additional scale it factorizes further
according to \cite{Chay:2003ju}
\be
  T_i^{\rm II} = H^{\rm II}_i * J ,
\ee
where the leading-order, ${\cal O}(\alpha_s)$, term is associated
with the tree approximation to both the hard kernels $ H^{\rm II}_i$
and the hard-collinear kernel (``jet function'') $J$.
In a previous paper~\cite{Beneke:2005vv} we computed the 
1-loop corrections  to the hard spectator-scattering
kernel $H_i^{\rm II}$  for the (topological) ``tree amplitudes'' in 
two-body decays. Here we extend this computation to the case of the
(topological) penguin amplitudes except for certain flavour-singlet 
terms that contribute only when $M_2$ is an $\eta$ or 
$\eta^\prime$ meson. Since $J$ is also known at one 
loop~\cite{Hill:2004if, Beneke:2005gs}, our result completes the set 
of spectator-scattering kernels at $\cO(\alpha_s^2)$.

The penguin amplitudes considered here provide the primary decay 
mechanism for $b\to s$ transitions. Their magnitudes and phases  
determine the size of the (direct) CP asymmetries in all charmless 
hadronic $B$ decays, since the necessary interference of decay 
amplitudes carrying different weak
and strong phases always involves a penguin amplitude. 
In this respect it is worth noting that 
the strong phases are confined to $T_i^{\rm I}$ at order $\alpha_s$. 
At $\cO(\alpha_s^2)$ a new source of strong phases appears in the
spectator-scattering kernels $T_i^{\rm II}$ via the one-loop 
correction to $H_i^{\rm II}$.
In~\cite{Beneke:2005vv} we found for the case of the topological
tree amplitudes that this contribution is
comparable in size to its $\cO(\alpha_s)$ counterpart in $T_i^{\rm I}$. 
Consequently it can change qualitatively the picture of CP asymmetries, 
and may be important both in accounting for experimental data and 
in predictions for yet unobserved asymmetries, as well as for 
disentangling possible new physics contributions
from the standard model background. In the case
of penguin amplitudes similarly large contributions could 
occur.\footnote{We note that numerically certain
$1/m_b$-suppressed, but ``chirally enhanced'' penguin 
amplitudes~\cite{Beneke:1999br} are important, which are currently 
known to $\cO(\alpha_s)$. The corresponding $\alpha_s^2$
contributions are not the subject of the present paper.}
Spectator-scattering corrections to one of the four penguin 
amplitudes have already been calculated in~\cite{Li:2005wx}. 
We discuss the difference between that calculation and ours
in Section~\ref{sec:calc}.

The organization of the paper is as follows. In
Section~\ref{sec:setup} we define the various flavour 
penguin amplitudes and discuss the diagram topologies that 
have to be calculated. We also set up the matching equations for the 
operators in the weak Hamiltonian to 
soft-collinear effective theory (SCET) that define the 
full set of hard-scattering kernels $H_i^{\rm II}$. These 
kernels are in turn expressed in terms of a complete set 
of ``primitive'' 1-loop hard-spectator-scattering kernels. 
Some details of their computation and the results for 
the primitive kernels are given in 
Section~\ref{sec:calc}. In Sections~\ref{sec:pengamps} and 
\ref{sec:ratios} we obtain 
the numerical values of the corrections to the penguin 
amplitudes $\alpha_{i}(M_1 M_2)$, 
$\alpha_{i,\rm EW}(M_1 M_2)$ ($i=3,4$), and provide updated 
results for some phenomenologically important 
penguin-to-tree and electroweak penguin-to-tree amplitude ratios. 
We conclude in Section~\ref{sec:conclude}.

\section{Structure of the penguin kernels}
\label{sec:setup}

\subsection{Flavour amplitudes}
\label{topoamp}

Our goal is to evaluate matrix elements of the
the effective weak Hamiltonian for $b\to D$ transitions 
given by~(see \cite{Beneke:2001ev}, where 
also numerical values of the Wilson coefficients $C_i$ 
are given)
\bea             
\cH_\eff &=& \frac{G_F}{\sqrt{2}} \sum_{p=u,c} \,V_{pD}^* V_{pb}
           \left( C_1 Q_1^p + C_2 Q_2^p
                + \sum_{i=3\dots10,7\gamma,8g} C_i Q_i \right) + \mathrm{h.c.},
\nonumber\\[0.4cm]
        && \hspace{-2cm} 
           Q_1^p = (\bar p_a b_a)_{V-A} (\bar D_b u_b)_{V-A} ,
           \qquad\quad 
           Q_2^p = (\bar p_b b_a)_{V-A} (\bar D_a u_b)_{V-A} ,
\nonumber\\[0.2cm]
        && Q_{3,5} = (\bar D_a b_a)_{V-A}
                \sum_q (\bar q_b q_b)_{V\mp A},
\nonumber\\
        && Q_{4,6} = (\bar D_b b_a)_{V-A}
                \sum_q (\bar q_a q_b)_{V\mp A},
\nonumber\\
        && Q_{7,9} = (\bar D_a b_a)_{V-A}
                \sum_q \frac{3}{2} e_q (\bar q_b q_b)_{V\pm A},
\nonumber\\
        && Q_{8,10} = (\bar D_b b_a)_{V-A}
                \sum_q \frac{3}{2} e_q (\bar q_a q_b)_{V\pm A},
\nonumber\\
        && \hspace{-2cm} 
        Q_{7\gamma} = -\frac{e m_b}{8\pi^2}\bar D \sigma_{\mu\nu} (1+\gamma_5)
                F^{\mu\nu} b,
 \qquad Q_{8g} = -\frac{g_s m_b}{8\pi^2}\bar D \sigma_{\mu\nu} (1 + \gamma_5)
                G^{\mu\nu} b,
\label{eq:weakham}
\eea
with $a,b$ denoting color, $D=d$ or $s$, and $(\bar q_1 q_2)_{V\mp A}
= \bar q_1\gamma_\mu (1\mp \gamma_5) q_2$. $e_q$ denotes the electric
charge of quark $q$ in units of the positron charge $e$,
and the sum over quarks extends over
$q=u,d,s,c,b$. The definition of the dipole operators corresponds to the
conventions $i D_\mu = i \partial_\mu + g_s G_\mu^A T^A + e e_q A_\mu$ and
$\sigma_{\mu\nu} = \frac{i}{2}[\gamma_\mu,\gamma_\nu]$. The effective 
Hamiltonian is understood to be renormalized in the NDR scheme as 
defined in \cite{Buras:1992tc}.

To organize the flavour quantum numbers of the factorized 
matrix elements (\ref{eq:qcdfact}) we match $\cH_\eff$ onto a 
transition operator ${\cal T}_A^p$ such 
that its matrix element is given by \cite{Beneke:2003zv}
\begin{equation}
\label{Top}
   \langle M_1'M_2'|{\cal H}_{\rm eff}|\bar B\rangle
   = \sum_{p=u,c} \lambda_p^{(D)}\,
   \langle M_1' M_2'|{\cal T}_A^p |\bar B\rangle.
\end{equation}
There are six different flavour structures required, resulting in
\begin{eqnarray}
\label{alphaidef}
   {\cal T}_A^p
   &=& \delta_{pu}\,\alpha_1(M_1 M_2)\,A([\bar q_s u][\bar u D])
    + \delta_{pu}\,\alpha_2(M_1 M_2)\,A([\bar q_s D][\bar u u])
    \nonumber\\[0.2cm]
   &&\mbox{}+ \alpha_3^p(M_1 M_2)\,\sum_q A([\bar q_s D][\bar q q])
    + \alpha_4^p(M_1 M_2)\,\sum_q A([\bar q_s q][\bar q D])
    \nonumber\\[-0.3cm]
   &&\mbox{}+ \alpha_{3,\rm EW}^p(M_1 M_2)\,\sum_q\frac32\,e_q\,
    A([\bar q_s D][\bar q q])
    \nonumber\\[-0.3cm]
   &&\mbox{}+ \alpha_{4,\rm EW}^p(M_1 M_2)\,\sum_q\frac32\,e_q\,
    A([\bar q_s q][\bar q D]), \qquad
\end{eqnarray}
where the sums now extend only over $q=u,d,s$, and $\bar q_s$ denotes the 
spectator anti-quark in the $\bar B$ meson. 
The coefficients $\alpha_i^p(M_1 M_2)$ 
contain all dynamical information, while the arguments of $A$ encode 
the flavour composition of the final state and hence determine the final
state to which a given term can contribute. The $\alpha_i$ parameters 
introduced in \cite{Beneke:2003zv} are in close correspondence with 
the widely used ``graphical'' or `` topological'' 
amplitudes~\cite{Chau:1990ay}: 
$\alpha_1$ ($\alpha_2$) with the colour-allowed (colour-suppressed) 
tree amplitude; $\alpha_4^p$ with the QCD penguin amplitude;  
$\alpha^p_3$ with the QCD flavour-singlet penguin amplitude; and 
$\alpha^p_{3,\rm EW}$ ($\alpha^p_{4,\rm EW}$) with the colour-allowed 
(colour-suppressed) electroweak penguin amplitude. We define
\begin{equation}\label{Adef}
   \langle M_1' M_2'|\alpha_i^p(M_1 M_2)\,A([\ldots][\ldots])
   |\bar B_{q_s}\rangle\equiv \alpha_i^p(M_1' M_2')\,A_{M_1' M_2'} 
\end{equation}
whenever the quark flavours of the first and second square bracket match 
those of $M_1'$ and $M_2'$, respectively, and  
\begin{equation}\label{Adef2}
   \langle M_1' M_2'|\alpha_i^p(M_1 M_2)\,A([\ldots][\ldots])
   |\bar B_{q_s}\rangle\equiv \alpha_i^p(M_2' M_1')\,A_{M_2' M_1'} 
\end{equation}
whenever the quark flavours of the first and second square bracket match 
those of $M_2'$ and $M_1'$. The quantity 
$A_{M_1 M_2}$ is given by
\begin{eqnarray}
A_{M_1 M_2} &=& i \frac{G_F}{\sqrt{2}} \left\{ \begin{array}{lcl}
   m_B^2 f_+^{BM_1}(0) f_{M_2} &\quad\quad& (M_1, M_2 = P) \\
   - m_B m_{V_1} (n_+ \cdot \epsilon^*_{M_1}) A_0^{BM_1}(0) f_{M_2} &&
                                             (M_1 = V, \,M_2=P) \\
   -m_B m_{V_2} (n_- \cdot \epsilon^*_{M_2}) f_+^{BM_1}(0) f_{M_2} &&
                                            (M_1 = P, \, M_2=V)\\
   m_{V_1} m_{V_2} (n_+ \cdot \epsilon^*_{M_1}) (n_- \cdot \epsilon^*_{M_2})
                         A_0^{BM_1}(0) f_{M_2} && (M_1, M_2 = V)
   \end{array} \right.
\label{AMM}
\end{eqnarray}
Here $f_{+}$ and $A_0$ denote pseudoscalar ($P$) and vector ($V$) meson 
form factors in the standard convention, and $f_{M_2}$ are the 
(longitudinal) decay constants. Here and in the remainder of the 
paper we consider only the longitudinal polarization state of the 
vector meson. This is sufficient for $B\to PV$ decays, but not for 
$B\to VV$, where two further transverse amplitudes are required for 
a complete description. The transverse amplitudes are, however,  
$1/m_b$-suppressed except for a certain electromagnetic 
amplitude \cite{Beneke:2005we}. (In (\ref{AMM}) $m_{V_1} 
(n_+ \cdot \epsilon^*_{M_1})$ is $\cO(m_B)$, hence all four 
expressions are of the same order in the heavy-quark expansion.)

In order to exemplify the notation consider 
the decay $\bar B_d\to \pi^0\rho^0$, for which $q_s=d$ and $D=d$. The 
spectator quark can go to either one of the two mesons, so $M_1$ can be 
$\pi^0$ or $\rho^0$. Hence, e.g.\
\begin{equation}
   \langle\pi^0\rho^0|\alpha_4^p(M_1 M_2) \sum_q A([\bar d q][\bar q d])
   |\bar B_d\rangle = \frac12 \left[ 
   \alpha_4^p(\pi^0\rho^0)\,A_{\pi^0\rho^0}
   + \alpha_4^p(\rho^0\pi^0)\,A_{\rho^0\pi^0} \right] .
\label{example}
\end{equation}
On the other hand, 
\begin{equation}
   \langle\pi^0\rho^0|\alpha_3^p(M_1 M_2) \sum_q A([\bar d d][\bar q q])
   |\bar B_d\rangle = 0,
\end{equation}
since $q=u,d$ contribute equally but with opposite sign for the mesons 
$\pi^0$ and $\rho^0$. We will assume isospin symmetry for the hadronic
parameters of the mesons. It is then conventional to express 
the form factors and decays constants through one representative  
member of the isospin multiplet, for instance of the charged pion 
in the case of pions. With this convention (\ref{Adef}) must be supplied 
with isospin Clebsch-Gordan coefficients $1,\pm 1/\sqrt{2}$ etc. from 
the flavour composition of the mesons. This is the origin of the
factor $1/2$ in (\ref{example}).

\subsection{Diagram topologies}

\begin{figure}[htp]
    \vspace{0.3cm}
\centerline{\includegraphics[width=15cm]{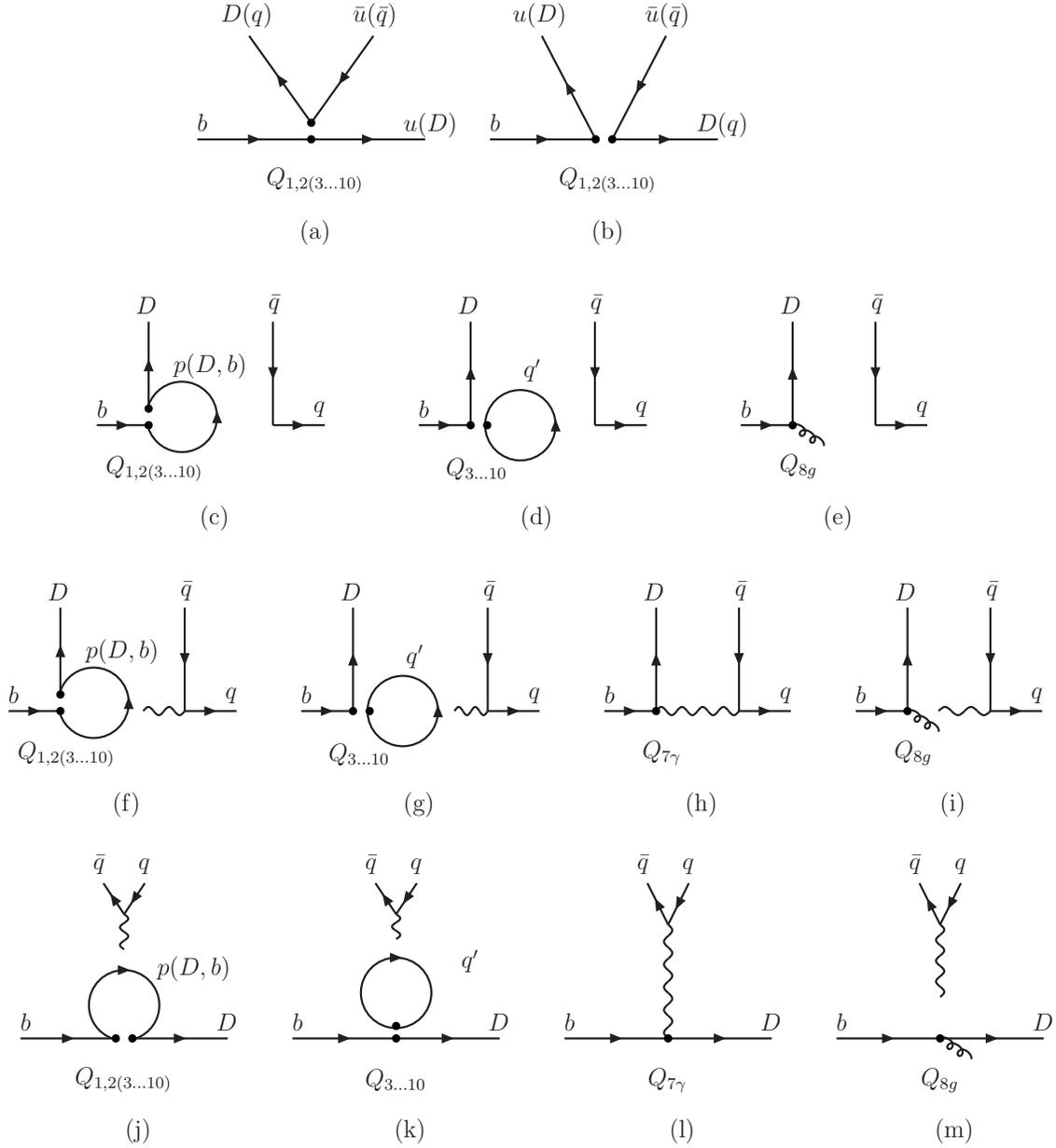}}
\caption{\label{fig1} The various
contractions of the external lines with the weak effective vertex,
excluding arbitrarily many additional gluons.
The connected fermion lines indicate the contraction 
of spinor indices. Terminology: (a) ``right insertion'',
(b) ``wrong insertion'', (c) [(f)] ``connected [photon] penguin'',
(d) [(g)] ``disconnected [photon] penguin'', 
(e) ``magnetic insertion''.
The photon can attach with its `loose' end anywhere except on
the line it originates from.
There are several more topologies (h) through
(m) contributing to electroweak penguin amplitudes (see text),
and some topologies (not shown) only contributing when at least
one final-state meson is an SU(3) flavour-singlet.
For a more detailed explanation, see the text. }
\end{figure}

Let us denote, in Figure~\ref{fig1} and below, the initial and final state
mesons by their flavour content. For instance, in 
Figure~\ref{fig1}(a) the $\bar B$ meson has the
flavour quantum numbers $[\bar q_s b]$, the final-state
meson $M_2$ represented by the up-going quark lines has
flavour content $[\bar u D]$, and the remaining meson $M_1$
$[\bar q_s u]$, where the $u$ line is drawn
to the right. (The spectator line is not shown in the figure.)

The flavour amplitudes
$\alpha_i^p$ in~\eq{alphaidef} receive perturbative contributions
due to several ways of contracting the quark lines in the
operators $Q_i$ with the valence quarks of the initial and
final states, and possibly with each other. Considering first the 
case where $M_2$ does not have a flavour-singlet component, 
i.e. disregarding $\alpha_3^p$ for the moment, the contractions 
are shown in Figure~\ref{fig1}. Here arbitrarily many gluons, that are
not shown, may connect the quark lines, or originate from a quark line and
connect with the spectator quark.
We also consider photon exchange to lowest order, as indicated in
Figure~\ref{fig1} (f) through (m). The last four contractions contribute
only if a vector meson is involved, due to the intermediate photon.

Focusing for the moment on the operator $Q_1^p$,
Figure~\ref{fig1}(a), which we call the ``right insertion'',
gives the contribution to the colour-allowed
tree amplitude $\alpha_1$ as can be seen by comparing the 
flavour labels of Figure~\ref{fig1}(a) with (\ref{alphaidef}).
Contraction (b), the ``wrong insertion'', gives the
contribution to the colour-suppressed tree
amplitude $\alpha_2$, while the ``connected penguin'' (c)
contraction contributes to the
(topological) penguin amplitude $\alpha_4^p$. 
For the operator $Q_2^p$, the roles of the ``right'' and ``wrong''
insertions are interchanged. The QCD penguin operators $Q_3 \dots Q_6$
contribute, through all of the contractions (b) (with relabeling 
$u\to D$, $\bar u\to \bar q$, and $D\to q$) through (d), to the
penguin amplitude $\alpha_4^p$ but not to $\alpha_{1,2}$. 
This is because of the sum over quarks
already present in the effective weak interaction operator. 
The magnetic penguin operator $Q_{8g}$ contributes
to $\alpha_4^p$ through contraction (e). In 
case of the electroweak penguin operators $Q_7 \dots Q_{10}$ 
the right insertions (a) contribute to $\alpha_{3, \rm EW}^p$, 
and the wrong insertions to $\alpha_{4, \rm EW}^p$. However, 
insertions of $Q_7 \dots Q_{10}$ into (c) and (d) 
do not contribute to the electroweak penguin amplitudes, 
but to electroweak corrections to the 
QCD penguin amplitudes, which we neglect. Thus, the second line of the
figure corresponds to contributions
to $\alpha_{4}^p$, while the first line can contribute
to any of the $\alpha_i^p$ coefficients depending on the operator 
that is inserted.

The flavour-singlet QCD penguin amplitude $\alpha_3^p$ is more 
complicated \cite{Beneke:2002jn}. There exists a contribution 
from the right insertion (a) which is similar to the ones discussed 
above, and which will be given below. In addition, however, there 
are penguin contractions, in which meson $M_2$ is made up only from 
gluons. The main reason for not calculating these contractions 
is that in case of the flavour-singlet penguin amplitude 
spectator scattering does not factorize in the 
form (\ref{eq:qcdfact}), but requires the introduction of 
a generalized non-local form factor \cite{Beneke:2002jn}.\footnote{
In soft-collinear effective theory this term is related to the 
SCET${}_{\rm I}$ operator with field content 
$[\bar \xi W_{c1} h_v][W_{c2}^\dagger iD_{c2}^{\mu_\perp} W_{c2}]$. 
This operator is relevant despite the colour-octet structure of 
the collinear-2 field product due to the non-decoupling 
of soft gluons at the level of power-suppressed  
interactions~\cite{Beneke:2002ph} in soft-collinear effective 
theory (SCET). The SCET rederivation of the 
QCD factorization formula for flavour-singlet mesons  
given in \cite{Williamson:2006hb} 
is not correct, because it neglects this contribution.}
Since the value of this form factor is unknown, the calculation 
of $\alpha_3^p$ is rather uncertain, and it is not useful to 
calculate loop corrections. 

\begin{figure}[t]
    \vspace{-2.8cm}
\centerline{\includegraphics[width=15cm]{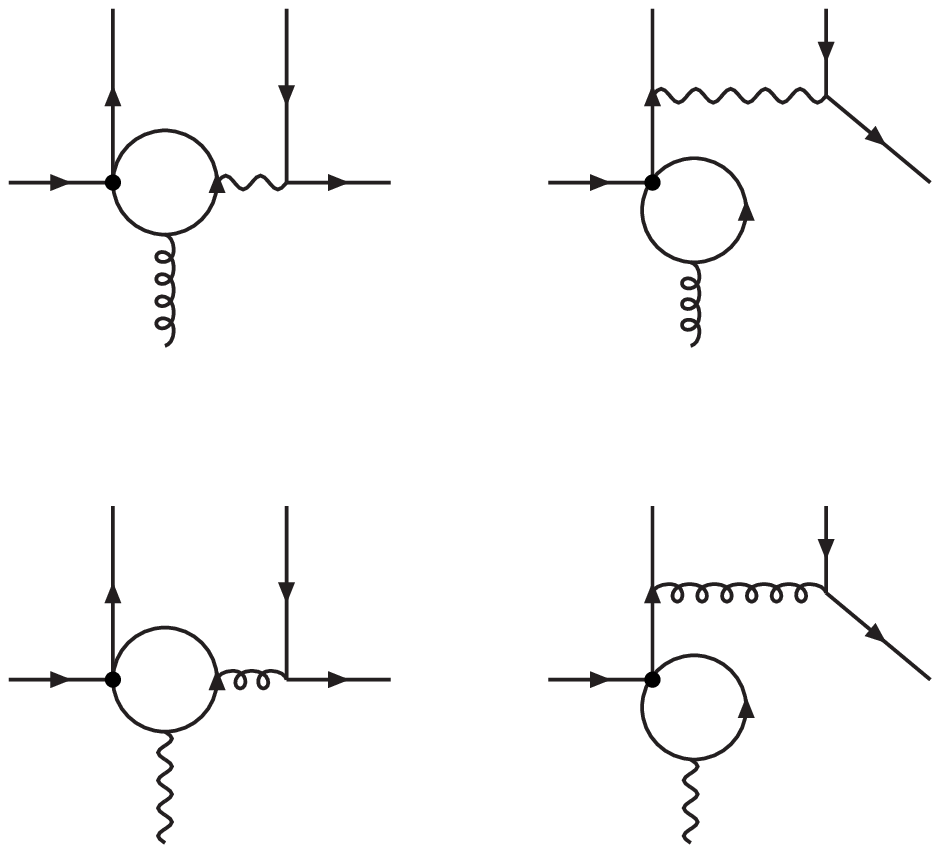}}
    \vspace*{-11.7cm}
\caption{\label{fig2} Examples of closely related 
diagrams that are interpreted as ``direct electroweak 
penguin'' contributions (first line), and as isospin-violating 
contributions (second line).}
\end{figure}

We also consider diagrams with photon exchange, but a 
complete calculation of QED corrections is beyond the scope of this 
paper. In fact, QED corrections to naive factorization have not 
even been considered for the simpler non-spectator contributions to the 
factorization formula up to now, except for an estimate of the 
soft photon contribution \cite{Baracchini:2005wp}. In general, 
electromagnetic corrections lead to isospin violation, incorporation
of which requires an extension of the 
parameterization (\ref{alphaidef}). Here, similar to 
\cite{Beneke:2001ev,Beneke:2003zv,Du:2000ff} for the non-spectator 
contributions, we restrict ourselves to photon exchange diagrams 
that directly give the charge and flavour structure of 
$\alpha^p_{3,\rm EW}$ and $\alpha^p_{4,\rm EW}$. 
These ``direct electroweak penguin'' contributions are shown in the 
third and fourth line of Figure~\ref{fig1}. When a photon is 
exchanged (besides, possibly, gluons) between the quark
loop and the $\bar q q$ quark line, a contribution from 
$Q_1 \ldots Q_6$ to the electroweak penguin amplitude 
$\alpha_{4,\rm EW}^p(M_1 M_2)$ 
arises for contractions (f) and (g), and to 
$\alpha_{3, \rm EW}^p(M_1 M_2)$ through contractions (j) and (k), 
if $M_2$ is a vector meson.  Finally, the magnetic dipole 
operators $Q_{7\gamma}$, $Q_{8g}$ contribute to
$\alpha_{4, \rm EW}^p$ (h,i) and, if $M_2=V$, to 
$\alpha_{3, \rm EW}^p$ via contractions (l) and (m). To make 
the distinction between ``direct electroweak penguin'' contributions 
and the remaining ones clearer we show in the first line of 
Figure~\ref{fig2} two diagrams that are included in our 
calculation of $\alpha^p_{4,\rm EW}$. On the other hand, the diagrams 
in the second line with the role of gluon and photon exchanged 
are not included, because they constitute isospin-violating contributions 
to the QCD penguin amplitude $\alpha^p_4$.

In summary, we see that there are a number of different 
short-distance coefficients, which in general receive contributions 
from several weak-interaction operators $Q_i$. The coefficients 
depend on (i) the Dirac structure of the operator $Q_i$, (ii)
its colour structure, (iii) the type of contraction,
(iv) for the penguin and magnetic contractions, whether a photon
is exchanged or not, and (v) on the mass of the quark in the fermion 
loop.

\subsection{Matching onto SCET${}_{\rm I}$}  
\label{sec:ffoview}

The hard-scattering kernels $T^{\rm I,II}_i$ can be determined by matching
QCD$\to$SCET${}_{\rm I}$ (resulting in $T_i^{\rm I}$, $H_i^{\rm II}$),
and in the case of $T^{\rm II}$ subsequently \sceti$\to$\scetii~ 
(resulting in $J$), whereby the kernels are identified with Wilson 
coefficients multiplying non-local operators in the effective
theories~\cite{Chay:2003ju,Beneke:2005vv,Beneke:2002ph,scet,BF03,Bauer:2002aj}.
The two steps are equivalent to extracting, respectively,
the hard and hard-collinear momentum regions
from  quark decay amplitudes according to the strategy of expanding 
Feynman diagrams by regions~\cite{Beneke:1997zp}. Single and
double logarithms of ratios of scales are summed
through renormalization group equations in SCET${}_{\rm I}$, and 
for the light-cone distribution amplitudes.
Already in SCET${}_{\rm I}$ modes corresponding to the two different
light-like directions decouple at leading power in $\Lambda/m_b$, 
except for the contribution to the flavour-singlet QCD penguin
amplitude $\alpha^p_3$ discussed in the previous subsection. Because
of this, in SCET${}_{\rm I}$ there are only right insertions, which moreover
factor into matrix elements of two currents.

Following the notations and conventions of \cite{Beneke:2002ph,BF03},
meson $M_1$, which picks up the spectator anti-quark from the $\bar B$
meson, moves in the direction of the light-like vector $n_-$.  
The collinear quark field for this direction is denoted by 
$\xi$ with $\slash{n}_-\xi=0$, the corresponding 
collinear gluon field is $A_{c1}$. 
The second meson $M_2$ moves in the opposite direction $n_+$, and 
the collinear fields for this direction are $\chi$, satisfying 
$\slash{n}_+\chi=0$, and $A_{c2}$. The heavy quark field $h_v$ 
is labeled by the time-like vector $v=(n_-+n_+)/2$ with $v^2=1$, 
and $\slash{v} h_v=h_v$. 

In~\cite{Beneke:2005vv} we argued that, ignoring colour and flavour, 
to leading power in $1/m_b$ only two operators in \sceti\ are needed 
to match the operators $Q_{1,2}^p$, as long as only current-current 
diagrams (contractions (a), (b) in Figure~\ref{fig1}) are considered.
The identification of these operators was based solely
on power-counting arguments in \sceti ~\cite{BF03}, except for the fact that
only left-chiral $n_+$-collinear quark fields $\chi$ were considered 
due to the $(V-A)\times(V-A)$ structure of $Q_{1,2}^p$.
Consequently, in addition to the extended flavour structures 
already discussed, the only novelty in the presence of the full set of
operators $Q_i$ and diagram topologies is the appearance of a second 
leading operator in the collinear-2 sector, of opposite chirality, given by 
$(\bar\chi \Wcp)(t n_-) \slash{n}_- (1 + \gamma_5) 
(\Wcp^\dag \chi )(0)$. Mixed-chirality bilinears built from the $\chi$ fields
necessarily involve either additional transverse derivatives or
a Dirac matrix carrying an uncontracted transverse Lorentz index.
In the former case they are power suppressed,
while in the latter case the resultant operator cannot contribute 
to pseudoscalar or longitudinally polarized vector mesons. 
Once again, the flavour-singlet penguin amplitude $\alpha^p_3$ is 
special and requires operators with $n_+$-collinear gluons. 
Thus, excepting $\alpha^p_3$, the four \sceti\ operators relevant 
to our calculation read
\bea
O_{L,R}^{\rm I}(t) &=& \Big[
  (\bar \chi\Wcp) (t n_-) \nmh (1 \mp \gamma_5) (\Wcp^\dag \chi) \Big]\;
        \Big[ \tilde C^{(A0)}_{f_+} \,(\xib W_{c1}) 
        \npslash (1 - \gamma_5) h_v 
\nonumber \\*
&&\hspace*{1cm} 
  - \frac{1}{m_b} \int d\hat{s}\, \tilde C^{(B1)}_{f_+}(\hat{s})\,
    (\bar \xi W_{c1}) \npslash [W^\dagger_{c1}i 
        \Slash{D}_{\perp c1} W_{c1}](s n_+)
                (1 + \gamma_5) h_v \Big] ,
\nonumber \\
O^{\rm II}_{L,R}(t,s) &=& \frac{1}{m_b} \Big[
   (\bar \chi\Wcp)(t n_-) \nmh (1 \mp \gamma_5) (\Wcp^\dag\chi)\Big] 
\nonumber \\ 
&&\hspace*{1cm} 
\times \Big[(\bar \xi W_{c1}) \nph [W^\dagger_{c1}i \Slash{D}_{\perp c1}
                 W_{c1}](s n_+) (1 + \gamma_5) h_v\Big],\quad
\label{ops}
\eea
where $O_L^{\rm I} = O^{\rm I}$ and $O_L^{\rm II} = O^{\rm II}$
in~\cite{Beneke:2005vv}.
The operators $O_{L,R}^{\rm I}$ include the short-distance coefficients 
$\tilde C^{(A0)}_{f_+}$, $\tilde C^{(B1)}_{f_+}(\hat{s})$ 
 \cite{Beneke:2005gs} such that 
their matrix elements are proportional to the form factor 
$f_+^{BM_1}(0)$ ($A_0^{B M_1}(0)$ for vector mesons) 
in QCD (not SCET${}_{\rm I}$).
As usual, in (\ref{ops}) fields without 
position argument are at $x=0$, and the field products within the 
large brackets are colour-singlets. When including 
electromagnetic interactions the ``direct electroweak 
penguins'' contribute to the coefficient functions of 
$O^{\rm I,II}_{L,R}$, but there are also SCET${}_{\rm I}$ 
operators of the above form with the transverse collinear 
gluon field $W^\dagger_{c1}i \Slash{D}_{\perp c1} W_{c1}$ 
replaced by a photon. This provides the sort of electromagnetic 
isospin-breaking, which as explained above we do not consider here. 
We can now take care of flavour by adding labels to the operators,
\be
O^{\rm I,II}_{L,R} \to
        O^{\rm I,II}_{L,R}([\bar q_s q_{M_1}] [\bar q_{M_2} q_{M_2}]) ,
\ee
where the labels $q_{M_{1,2}}$ and $\bar q_{M_2}$ give the flavours of
the fields $\bar \xi$, $\bar \chi$, and $\chi$, respectively, and
the redundant spectator label $\bar q_s$ has been added to match the 
notation (\ref{alphaidef}). Then up the $n_+$-collinear gluon operators 
mentioned above, whose matrix elements
contribute only to flavour-singlet $M_2$ (via a modification of
$\alpha^p_3$), at leading power in $1/m_b$ the complete weak
Hamiltonian~(\ref{eq:weakham}) can be accounted for in 
SCET${}_{\rm I}$ by 
\bea
\label{eq:matchweakham}
{\cal H_{\rm eff}} &=& \frac{G_F}{\sqrt{2}} \sum_{p=u,c} V^*_{pD}V_{pb} \Bigg(
    \delta_{pu} \,\Bigg\{ T_1^{\rm I} * O^{\rm I}_L([\bar q_s u] [\bar u D])
                    + H_1^{\rm II} * O^{\rm II}_L([\bar q_s u] [\bar u D])
\nonumber \\ && \qquad \quad \qquad \qquad
                     + T_2^{\rm I} * O^{\rm I}_L([\bar q_s D] [\bar u u])
                     + H_2^{\rm II} * O^{\rm II}_L([\bar q_s D] [\bar u u])
                \Bigg\}
\nonumber \\ && \qquad 
  + \sum_{k=L,R} \Bigg\{
        T_{3k}^{{\rm I},p} * \sum_q O^{\rm I}_k([\bar q_s D] [\bar q q])
      + H_{3k}^{{\rm II},p} * \sum_q O^{\rm II}_k([\bar q_s D] [\bar q q])
\nonumber \\ && \qquad \qquad 
      + T_{3k,\rm EW}^{{\rm I},p} * \sum_q \frac{3}{2} 
        e_q O^{\rm I}_k([\bar q_s D] [\bar q q])
      + H_{3k,\rm EW}^{{\rm II},p} * \sum_q \frac{3}{2} 
        e_q O^{\rm II}_k([\bar q_s D] [\bar q q]) \Bigg\} \Bigg)
\nonumber \\ && \qquad 
  + \sum_{k=L,R} \Bigg\{
        T_{4k}^{{\rm I},p} * \sum_q O^{\rm I}_k([\bar q_s q] [\bar q D])
      + H_{4k}^{{\rm II},p} * \sum_q O^{\rm II}_k([\bar q_s q] [\bar q D])
\\ && \qquad \qquad 
      + T_{4k,\rm EW}^{{\rm I},p} * \sum_q \frac{3}{2} 
        e_q O^{\rm I}_k([\bar q_s q] [\bar q D])
      + H_{4k,\rm EW}^{{\rm II},p} * \sum_q \frac{3}{2} 
        e_q O^{\rm II}_k([\bar q_s q] [\bar q D]) \Bigg\} \Bigg)
\nonumber            
\eea
where we employed the notation
\begin{equation}
T^{\rm I}_{ik} * O_k^{\rm I} = \int d\hat{t}\,\tilde{T}^{\rm I}_{ik}(\hat t) 
O_k^{\rm I}(t),
\qquad 
H_{ik}^{\rm II} * O_k^{\rm II} =
\int d\hat{t}d\hat{s}\,\tilde{H}_{ik}^{\rm II}(\hat t,\hat s)O_k^{\rm II}(t,s)
\end{equation}
with $\hat s=n_+ p^\prime s= m_B s$, $\hat t=n_- q \,t=m_B t$, 
and $p^\prime$ ($q$) the momentum of $M_1$ ($M_2$).
As in (\ref{alphaidef}),
the sums over $q$ extend only over the light quarks $u$, $d$, $s$, 
eventually implying that we neglect the ``intrinsic charm'' content of the 
mesons.
Of the various matching coefficients in~(\ref{eq:matchweakham}), the
$T_{ik}^{\rm I}(u) = \int d \hat t \,e^{i u \hat t}  
\,\tilde T_{ik}^{\rm I}(\hat t)$ are all known to the 1-loop order 
($\alpha_s$)~\cite{Beneke:1999br,Beneke:2001ev,Beneke:2003zv}.
The 1-loop ($\alpha_s^2$) corrections to
\begin{equation}
H^{\rm II}_{1,2}(u,v) = \int d\hat{t}d\hat{s}\,
e^{i (u\hat{t}+(1-v)\hat {s})}\,
\tilde{H}^{\rm II}_{1,2}(\hat t,\hat s)
\end{equation}
have been computed in~\cite{Beneke:2005vv}. In this paper we will compute the
remaining coefficients $H^{{\rm II},p}_{4k}$, $H^{{\rm II},p}_{3k,\rm
  EW}$, $H^{{\rm II},p}_{4k,\rm EW}$, and parts of $H^{{\rm
    II},p}_{3k}$.
We do not perform an expansion in $m_c/m_b$ in the matching 
calculation. The hard matching coefficients are therefore 
functions of the ratio $s_c=m_c^2/m_b^2$, whenever diagrams with 
internal charm-quark loops contribute.

The individual terms in~(\ref{eq:matchweakham}) are in close
correspondence with the $\alpha_i^p(M_1 M_2)$ amplitude parameters.
The precise connection follows by evaluating the matrix element of
(\ref{eq:matchweakham}). Because the SCET 
Lagrangian contains no leading-power interactions between 
the collinear-2 and collinear-1 fields, the matrix elements
of  $O^{\rm I}_k(t,s)$, $O^{\rm II}_k(t,s)$ fall apart into two factors each.
For a pseudoscalar $M_2=P$,
\begin{eqnarray}
&& \langle P|(\bar \chi\Wcp)(t n_-) \nmh (1 \pm \gamma_5) (\Wcp^\dag\chi)
  |0\rangle = \mp \frac{i f_{P} m_B}{2} 
  \int_0^1 du\,e^{i u \hat{t}} \,\phi_{P}(u),
\end{eqnarray}
while for a vector $M_2=V$ with polarization vector $\epsilon_\mu$ we have
\begin{eqnarray}
&& \langle V|(\bar \chi\Wcp)(t n_-) \nmh (1 \pm \gamma_5) (\Wcp^\dag\chi)
  |0\rangle = - \frac{i f_{V} m_V}{2} \,n_- \cdot \epsilon^*
  \int_0^1 du\,e^{i u \hat{t}} \,\phi_{V}(u) ,
\end{eqnarray}
such that only the longitudinal polarization state contributes. 
(The apparent suppression $m_V/m_B$ is cancelled
by the polarization vector.) Here $\phi_P(u)$ ($\phi_V(u)$) denotes the 
leading-twist light-cone distribution amplitude of a pseudoscalar 
(longitudinally polarized vector) meson. 
Using~\cite{Beneke:2005gs}
\begin{eqnarray} 
\label{me1}
&& \!\langle P|(\bar \xi W_{c1}) \nph 
  [W^\dagger_{c1}i \Slash{D}_{\perp c}W_{c1}](s n_+)
  (1 + \gamma_5) h_v|\bar B\rangle = -m_b m_B \int_0^1 d\tau 
  \,e^{i\tau\hat{s}}\,\Xi_{P}(\tau),
\\
&& \!\langle V|(\bar \xi W_{c1}) \nph 
  [W^\dagger_{c1}i \Slash{D}_{\perp c}W_{c1}](s n_+)
  (1 + \gamma_5) h_v|\bar B\rangle = m_b m_V (n_+\cdot\epsilon^*)
   \!\int_0^1 d\tau 
  \,e^{i\tau\hat{s}}\,\frac{m_B}{2 m_V}\Xi_{\parallel}(\tau),
\nonumber
\end{eqnarray}
and defining 
\be
\hat\Xi_{M_1}(\tau)=\left\{\begin{array}{lll}
\Xi_{P}(\tau) &\qquad& M_1=P \\[0.2cm]
\displaystyle \frac{m_B}{2 m_V}\Xi_{\parallel}(\tau) &\qquad&M_1=V
\end{array}
\right.
\ee
we obtain
\begin{eqnarray}
   \alpha_{ik}^p(M_1 M_2) &=&  \int_0^1 du \,T_{ik}^{\rm I}(u)
   \phi_{M_2}(u)
   \nonumber\\ 
        &&- \,\frac{1}{2 F^{B M_1}(0)} 
        \int_0^1 du dv \,H_{ik}^{{\rm II},p}(u,v) 
        \,\hat\Xi_{M_1}(1-v) \phi_{M_2}(u)
        \label{eq:alphadef}
\end{eqnarray}
with $F^{B M_1}(0)=f_+^{B M_1}(0)$ when $M_1=P$, and 
$F^{B M_1}(0)=A_0^{B M_1}(0)$  
for $M_1=V$. The index $i$ applies to $i=1,2,3,4,3\mbox{EW},
4\mbox{EW}$, and $k=L,R$ except for $i=1,2$, where $k$ is empty (as is $p$). 
The two coefficients 
$\alpha_1(M_1 M_2)$, $\alpha_2(M_1 M_2)$ correspond to the 
``tree'' flavour-amplitudes in (\ref{alphaidef}), while the 
four ``penguin'' amplitudes are given by
 \begin{eqnarray}
\alpha^p_3(M_1 M_2) &=& \alpha_{3L}^p(M_1 M_2) \mp 
  \alpha_{3R}^p(M_1 M_2), 
\nonumber\\
\alpha^p_4(M_1 M_2) &=& \alpha_{4L}^p(M_1 M_2) \mp 
  \alpha_{4R}^p(M_1 M_2) ,
\nonumber\\
\alpha^p_{3,\rm EW}(M_1 M_2) &=& \alpha_{3L,\rm EW}^p(M_1 M_2) \mp 
  \alpha_{3R,\rm EW}^p(M_1 M_2), 
\nonumber\\
\alpha^p_{4,\rm EW}(M_1 M_2) &=& \alpha_{4L,\rm EW}^p(M_1 M_2) \mp 
  \alpha_{4R,\rm EW}^p(M_1 M_2) .
\end{eqnarray}
Here the upper (lower) signs  
correspond to the case $M_2=P(V)$.\footnote{In the older $a_i$ notation 
the coefficients $\alpha^p_{3,4,3\rm EW,4\rm EW}$ 
of the left-handed operators 
are related to $a_{3,4,9,10}$, the right-handed $\alpha^p_{3,3\rm EW}$
to $a_{5,7}$. The remaining 
$\alpha^p_{4R}$, $\alpha^p_{4R,\rm EW}$ vanish. The coefficients 
$a_{6},\,a_8$ in the $a_i$ notation correspond to power-suppressed 
penguin amplitudes. Because they have the same flavour structure, 
they are included in the definition of $\alpha_4$ and 
$\alpha_{4,\rm EW}$, respectively, in \cite{Beneke:2003zv}. 
Similarly, the contributions from gluon operators omitted above
are included in $\alpha_3^p$.
This will be assumed 
in the following. See also (\ref{apar}) 
below. } The generalized form factors 
$\hat \Xi(\tau)$ factorize into light-cone distribution 
amplitudes after matching them to SCET${}_{\rm II}$ \cite{BF03}. 
The result is \cite{Beneke:2005gs}
\be
\hat\Xi(\tau) = \frac{m_B}{4m_b}\int_0^\infty
\frac{d\omega}{\omega}\,\hat{f}_B\phi_{B+}(\omega)
\int_0^1 dw \,f_{M_1}\phi_{M_1}(w)\,J(\tau;w,\omega)
\label{Xifact}
\ee
with $J=J_\parallel$, the hard-collinear kernel, known to 
order $\alpha_s^2$~\cite{Hill:2004if, Beneke:2005gs}, and 
$\hat{f}_B$ the static $B$ meson decay constant as 
defined in~\cite{Beneke:2005gs}.

\subsection{Primitive kernels}

\begin{table}
\begin{center}
\begin{tabular}{|c|ccccc|}
\hline
Structure & \multicolumn{5}{c|}{Contraction} \\
          & (a) & (b) & (c) & (d) & (e) \\
\hline
&&&\phantom{$s_{3,\rm EW},s_{4,\rm EW}$}&
\phantom{$s_{3,\rm EW},s_{4,\rm EW}$}& \\[-0.5cm]
$P_{LL}$ &      $r_2,r_1$ & $r_1,r_2$ & $s_3, s_4$ & $s_1,s_2$ & ---\\
$P_{LR}$ &      $r_4,r_3$ & $\star$ & $s_3', s_4'$ & $s_1', s_2'$ &
--- \\
$P_{\rm mag}$ & --- & --- & --- & --- & $m_1$ \\
\hline
          & (f) & (g) & (h) & (i) &  \\
\hline
$P_{LL}$ &  $s_{3,\rm EW},s_{4,\rm EW}$ & $s_{1,\rm EW}, s_{2,\rm
  EW}$ &  --- & --- & \\
$P_{LR}$ &  $s_{3,\rm EW}',s_{4,\rm EW}'$ & $s_{1,\rm EW}', s_{2,\rm EW}'$
& --- & --- & \\
$P_{\rm mag}$ & --- & --- &  $m_{1, \rm EW}$ & $m_{2, \rm EW}$  & \\
\hline
          & (j) & (k) & (l) & (m) &  \\
\hline
$P_{LL}$ &  $s_{5,\rm EW}, [s_{6,\rm EW}]$ & $[s_{7,\rm EW}], s_{8,\rm
  EW}$ &  --- & --- & \\
$P_{LR}$ &  $s_{5,\rm EW}',s_{6,\rm EW}'$ & $[s_{7,\rm EW}'], s_{8,\rm EW}'$
& --- & --- & \\
$P_{\rm mag}$ & --- & --- &  $m_{3, \rm EW}$ & $m_{4, \rm EW}$  & \\
\hline
\end{tabular}
\end{center}
\caption{Definition of the primitive kernels. Where two kernels are given,
the first corresponds to the colour structure of $Q_1$, the second
to that of $Q_2$. The ``star'' means that there are no corresponding
kernels at leading power. Kernels in parentheses vanish at the 1-loop
order.
\label{tab:primitive}}
\end{table}

As explained at the end of Section~\ref{topoamp}, the short-distance
coefficients encountered in matching an operator $Q_i$ in~(\ref{eq:weakham})
can only depend on the Dirac structure, the colour structure, the
type of contraction, the mass of the quark in penguin loops as
  far as present,
 and in the case of penguin contractions and of
$Q_{8g}$, whether a photon is exchanged with the quark line on the right
or not. Stripping the operators of all light-flavour and colour labels and
an overall normalization factor, we encounter the full-QCD Dirac
structures
\bea
        P_{LL}  &=&     [\bar q \gamma^\mu (1-\gamma_5) q]
                           [\bar q \gamma_\mu (1-\gamma_5) b],
\nonumber \\
        P_{RL}  &=&     [\bar q \gamma^\mu (1+\gamma_5) q]
                           [\bar q \gamma_\mu (1-\gamma_5) b],
\nonumber \\
        P_{\rm mag} &=& -\frac{g m_b}{8\pi^2} \,\bar q 
                        \sigma_{\mu\nu} (1+\gamma_5) b,
\eea
where $q$ denotes some light quark field (no summation implied) and
$g = e$ or $g_s$. Their insertions into the contractions in
Figure~\ref{fig1} define primitive kernels according to 
Table~\ref{tab:primitive}.

Of all these kernels, $r_1$ and $r_2$ have been given
in~\cite{Beneke:2005vv}.\footnote{The calculation of $r_{1,2}$ 
has been repeated in \cite{Kivel:2006xc}. There is a difference 
in the results for $r_1$, which is simply the colour factor 
$(C_F-C_A/2)$ times the tree-level kernel. Such a difference is 
likely to originate from an inconsistent treatment of ultraviolet 
or infrared singularities. While \cite{Beneke:2005vv} uses 
dimensional regularization for infrared singularities, and therefore 
has to deal with evanescent operators, off-shell IR regularization 
is employed in \cite{Kivel:2006xc}, the price for which is the 
calculation of non-trivial SCET${}_{\rm I}$ matrix elements. If no 
error is made, the result for the hard-scattering kernels should 
be the same in both methods. In addition, in \cite{Kivel:2006xc} 
a projection on the Dirac structure of the SCET${}_{\rm I}$ operator 
is taken. It is not obvious that such projections commute 
with $\overline{\rm MS}$ renormalization, and in general 
they do not.
{\bf Note added:} After the submission of the present paper for
publication,
a corrected version of \cite{Kivel:2006xc} appeared, which now agrees
with the results of~\cite{Beneke:2005vv}.} The fact that these are sufficient 
to account for both ``right'' and ``wrong''
insertions (a) and (b) can be traced to Fierz symmetry. This is only valid
because of the Fierz properties of the scheme used to define the renormalized
effective weak Hamiltonian. In the case of penguin contractions, we will 
find, for instance, $s_3 \neq s_2$, although the 
differences between such naively Fierz related kernels always 
assume a very simple form. In the present paper we also calculate 
the insertion (a) of the $P_{RL}$ structure. 
Insertion (b) results in a $1/m_b$ correction (related to the infamous 
power-suppressed but ``chirally enhanced'' scalar 
penguin amplitudes), so only two primitive kernels $r_{3,4}$ 
are needed. On the other hand, the
fact that the penguin contractions (c) and (d) contribute for both
$V-A\times V-A$ and $V-A\times V+A$ operators and for both 
colour structures requires the introduction of eight  
kernels $s_i, s_i^\prime$, $i=1\ldots 4$. The remaining 
kernels are related to electroweak penguin terms and the 
insertions of the magnetic dipole operators. 

We can now express the hard-scattering kernels 
$H_{ik}^{\rm II}$ in~(\ref{eq:matchweakham}) in terms of these
primitive building blocks as ($\bar n_f = n_f-2=3$ the number of flavours 
treated as massless)
\bea
  H_{4L}^{{\rm II},p} &=& \frac{2}{N_c} \biggl\{
    \Bigl[ \frac{1}{\bar u} + \frac{\alpha_s}{4\pi} r_1 \Bigr] C_3
        + \frac{\alpha_s}{4\pi} r_2 C_4
\nonumber\\ &&
        + \frac{\alpha_s}{4\pi} \Bigl[C_1 s_3(s_p) + C_2 s_4(s_p)
          + C_3 \Big(s_3(0) + s_3(1)\Big) + C_4 \Big(s_4(0) + s_4(1)\Big)
\nonumber\\ && \qquad
          + \,C_5 s_3'(1) + C_6 s_4'(1)
\nonumber\\ && \qquad
          + \Big(s_1(1) + s_1(s_c) + \bar n_f s_1(0) \Big) C_3
          + \Big(s_1'(1) + s_1'(s_c) + \bar n_f s_1'(0) \Big) C_5
\nonumber\\ && \qquad
          + \Big(s_2(1) + s_2(s_c) + \bar n_f s_2(0) \Big) C_4
          + \Big(s_2'(1) + s_2'(s_c) + \bar n_f s_2'(0) \Big) C_6
        \Bigr] 
\nonumber\\ && \qquad
 + \,m_1 C_{8g}
\biggr\}  ,
\label{H4L}
\\
  H_{4R}^{{\rm II},p} &=& 0 ,
\\[0.2cm]
  H_{3L}^{{\rm II},p} &=& \frac{2}{N_c} \biggl\{
    \Bigl[ \frac{1}{\ub} + \frac{\alpha_s}{4\pi} r_1 \Bigr] C_4
        + \frac{\alpha_s}{4\pi} r_2 C_3 \biggr\} ,
\\[0.2cm]
  H_{3R}^{{\rm II},p} &=& \frac{2}{N_c} \biggl\{
    \Bigl[ -\frac{1}{u} + \frac{\alpha_s}{4\pi} r_3 \Bigr] C_6
        + \frac{\alpha_s}{4\pi} r_4 C_5 \biggr\} ,
\\[0.2cm]
  H_{4L,\rm EW}^{{\rm II},p} &=& \frac{2}{N_c} \biggl\{
        \Bigl[ \frac{1}{\bar u} + \frac{\alpha_s}{4\pi} r_1 \Bigr] C_9
        + \frac{\alpha_s}{4\pi} r_2 C_{10}
\nonumber\\ &&
        +\frac{2}{3} \frac{\alpha_\mathrm{em}}{4\pi} 
        \Bigl[ e_u C_1 s_{3,\rm EW}(s_p) + e_u C_2 s_{4,\rm EW}(s_p)
        + e_d C_3 \Big(s_{3,\rm EW}(0) + s_{3,\rm EW}(1)\Big)
\nonumber\\ && \qquad
        + \,e_d C_4 \Big(s_{4,\rm EW}(0) + s_{4,\rm EW}(1)\Big)
        + e_d C_5 s_{3,\rm EW}'(1) + e_d C_6 s_{4,\rm EW}'(1)
\nonumber\\ && \qquad
        + \Big( e_d s_{1,\rm EW}(1) + e_u s_{1,\rm EW}(s_c) \Big) C_3
        + \Big( e_d s_{2,\rm EW}(1) + e_u s_{2,\rm EW}(s_c) \Big) C_4
\nonumber\\ && \qquad
        + \Big( e_d s_{1,\rm EW}'(1) + e_u s_{1,\rm EW}'(s_c) \Big) C_5
        + \Big( e_d s_{2,\rm EW}'(1) + e_u s_{2,\rm EW}'(s_c) \Big) C_6
\nonumber\\ && \qquad
        + \, m_{1, \rm EW} C_{7\gamma}^{\rm eff} 
        + m_{2, \rm EW} C_{8g}^{\rm eff}
        \Bigr] \biggr\},
\\
  H_{4R,\rm EW}^{{\rm II},p} &=& 0 ,
\\[0.2cm]
  H_{3L,\rm EW}^{{\rm II},p} &=& \frac{2}{N_c} \biggl\{
    \Bigl[ \frac{1}{\ub} + \frac{\alpha_s}{4\pi} r_1 \Bigr] C_{10}
        + \frac{\alpha_s}{4\pi} r_2 C_9
\nonumber\\ && 
        + \frac{2}{3} \frac{\alpha_{\rm em}}{4\pi} \Bigl[
                e_u s_{5, \rm EW}(s_p) C_1
                + e_d (s_{5,\rm EW}(0) + s_{5, \rm EW}(1)) C_3
\nonumber\\ && \qquad
        + \Big( e_d s_{8,\rm EW}(1) + e_u s_{8,\rm EW}(s_c) \Big) C_4
        + \Big( e_d s_{8,\rm EW}'(1) + e_u s_{8,\rm EW}'(s_c) \Big) C_6
\nonumber\\ && \qquad
        + \,m_{3,\rm EW} C_{7\gamma}^{\rm eff}
        + m_{4, \rm EW} C_{8g}^{\rm eff}
        \Bigr]\biggr\} ,
\\[0.2cm]
  H_{3R,\rm EW}^{{\rm II},p} &=& \frac{2}{N_c} \biggl\{
    \Bigl[ -\frac{1}{u} + \frac{\alpha_s}{4\pi} r_3 \Bigr] C_8
        + \frac{\alpha_s}{4\pi} r_4\, C_7
\nonumber\\ && 
        + \frac{2}{3} \frac{\alpha_{\rm em}}{4\pi} \Bigl[
                e_u s_{5, \rm EW}(s_p) C_1
                + e_d (s_{5,\rm EW}(0) + s_{5, \rm EW}(1)) C_3
\nonumber\\ && \qquad
        + \Big( e_d s_{8,\rm EW}(1) + e_u s_{8,\rm EW}(s_c) \Big) C_4
        + \Big( e_d s_{8,\rm EW}'(1) + e_u s_{8,\rm EW}'(s_c) \Big) C_6
\nonumber\\ && \qquad
        + \, m_{3,\rm EW} C_{7\gamma}^{\rm eff}
        + m_{4, \rm EW} C_{8g}^{\rm eff}
        \Bigr]\biggr\} .
\label{H3REW}
\eea
Here the argument of the primitive kernels denotes the quark mass 
ratio $s_q=m_q^2/m_b^2$ with $q$ the quark flavour in the fermion 
loop of the penguin contraction. We also used $2 e_d+e_u=0$ 
to simplify some of the electroweak penguin kernels, 
and introduced the standard ``effective'' Wilson coefficients, 
$C_{7\gamma}^{\rm eff}=C_{7\gamma}+ e_d (C_5+N_c C_6)$, 
$ C_{8g}^{\rm eff}= C_{8g}+C_5$ to combine some ultraviolet 
contributions from fermion loops with the magnetic penguin 
coefficients. This convention modifies the primed 
primitive kernels such that $s_{5, \rm EW}'$, $s_{6, \rm EW}'$ vanish 
at one loop.  For $H_{3L,\rm EW}^{{\rm II},p}$ and 
$H_{3R,\rm EW}^{{\rm II},p}$, the terms on the last three lines of
each that stem from the insertions (j) to (l) in 
Figure~\ref{fig1} are identical, so they cancel
out for $M_2=P$ as they should. To simplify the notation for 
$H_{3L,\rm EW}^{{\rm II},p}$ and $H_{3R,\rm EW}^{{\rm II},p}$, 
we already dropped the primitive kernels that vanish at 
one loop. 

\section{Computation and results}
\label{sec:calc}

In this section we provide some technical remarks on the calculation 
and then summarize the results for the primitive kernels. We calculate
the quark-gluon amplitude 
$b(p) \to q(q_1) \bar q(q_2)  q(p^\prime_1) g(p^\prime_2)$ 
with insertions of $Q_i$ in QCD at 1-loop and subtract the
corresponding SCET${}_{\rm I}$ matrix elements to obtain the 
hard-scattering coefficient. Insertions of the magnetic dipole 
operators require only a tree-level calculation. At leading order 
in the $1/m_b$ expansion it is almost always possible to 
approximate the external momenta by their leading components
$p=m_b v$, $q_1=u m_b n_+/2$, $q_2=\bar u m_b n_+/2$, 
$p^\prime_1=v m_b n_-/2$, $p^\prime_2=\bar v m_b n_-/2$, 
which makes the calculation particularly simple in dimensional 
regularization. Exceptions to this are provided by diagrams 
with intermediate lines whose virtuality vanishes in this 
approximation, such as (a6), (e4) and all the diagrams in the third
line of Figure~\ref{fig3}. 

\subsection{Technical details}

\subsubsection*{Vertex contractions (Figure~\ref{fig1} (a),(b))}

The calculation of the new kernels $r_{3,4}$ related to insertions 
of $(V-A)\times(V+A)$ operators proceeds analogously 
to the one of $r_{1,2}$ described in detail 
in~\cite{Beneke:2005vv}. Only the right insertions need to be 
computed, since the wrong insertions match to power-suppressed 
SCET${}_{\rm I}$ operators. Substituting $1-\gamma_5 \to 
1+\gamma_5$ in Eq.~(17) of~\cite{Beneke:2005vv} to deal 
with the new $O^{\rm II}_R$ operators leads to a SCET operator 
basis, in which $O_{2-4}$ are no longer evanescent (vanishing 
in four dimensions). We can remedy this by choosing a new 
basis, in which all $\gamma^\mu_\perp$ stand to the right 
of all other transverse Dirac matrices. The treatment 
of IR singularities can then be done as in~\cite{Beneke:2005vv}. 
In the calculation of the matrix element of the evanescent 
operator $O_2$, one now encounters in addition to the 
contribution shown in Figure~4 of~\cite{Beneke:2005vv} 
a contribution related to the matrix element of the 
$\bar \xi A_{\perp c1} h_v$ part of the operator. This contribution 
can be deduced from the ultraviolet pole of $B$ in 
Eq.~(39) of~\cite{Beneke:2005gs}.

\subsubsection*{QCD penguin contractions 
(Figure~\ref{fig1} (c) -- (e))}

\begin{figure}[t]
    \vspace{-1.8cm}
    \hspace*{-0.3cm}
\centerline{\includegraphics[width=12cm]{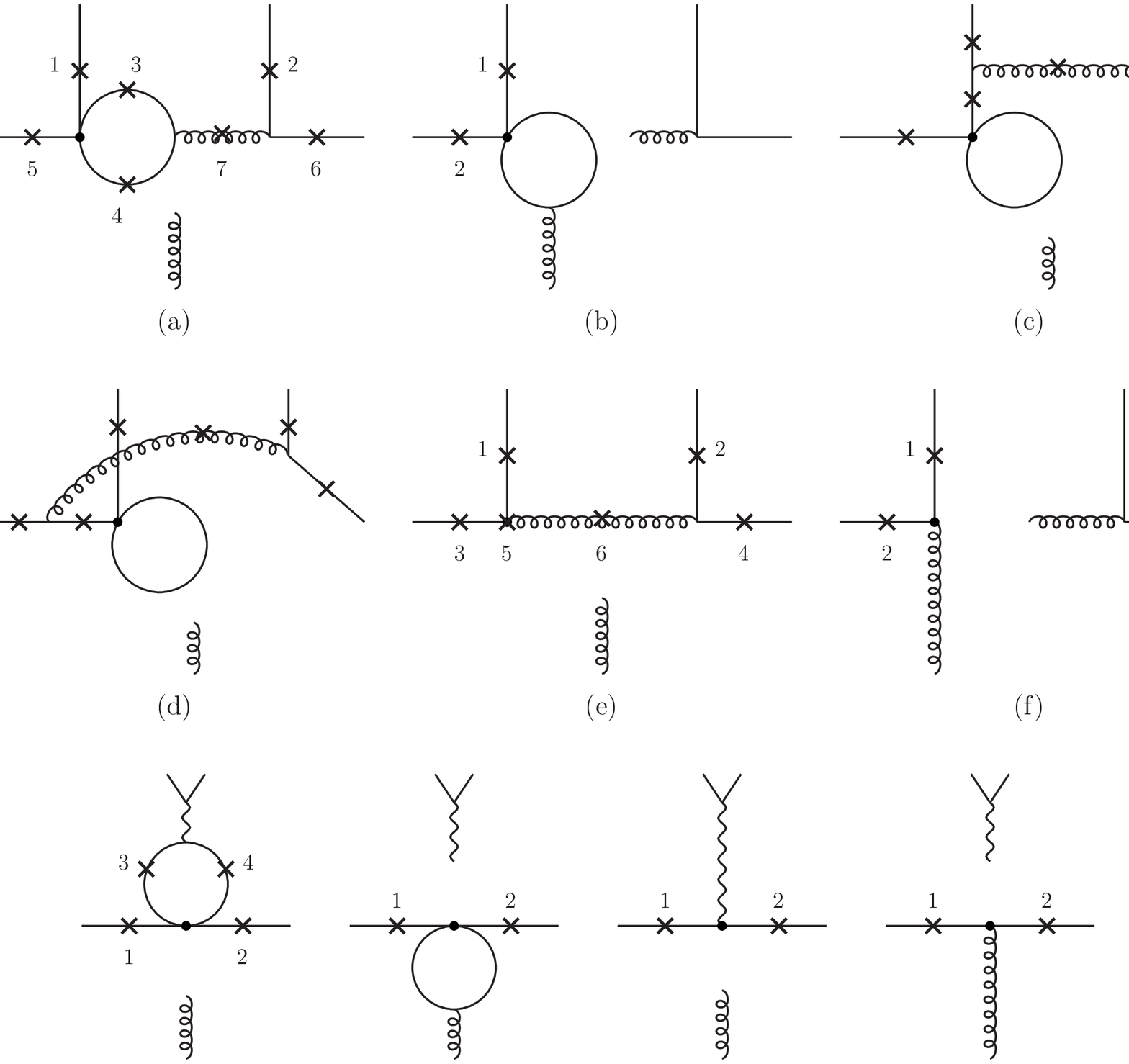}}
    \vspace*{-3.4cm}
\caption{\label{fig3} Summary of penguin-contraction diagrams. 
The `loose' gluon or photon can attach to the crosses. Corresponding 
to diagram classes (a)-(f) there exist the diagrams with the 
horizontal gluon line replaced by a photon line. (For these 
(a7), (e5) and (e6) are absent.)}
\end{figure}

The penguin contraction diagrams are shown in Figure~\ref{fig3}, 
of which the first two lines are relevant to the QCD penguin 
amplitude $\alpha^p_4$. The calculation of these diagrams is much 
simpler than the calculation of the vertex contractions. In
particular, they are infrared finite, and no evanescent operators
in SCET${}_{\rm I}$
need to be considered. As in the case of the vertex contractions, 
one must take into account that the QCD factorization formula 
uses full-QCD form factors by convention, hence the second line 
in the definition of $Q^I_{L,R}$ in (\ref{ops}). As a consequence, 
the ``form-factor subtraction'' 
\be
2 T^{\rm I(1)}_{ik} C^{(B1)(0)}_{f_+} = - 2 T^{\rm I(1)}_{ik}
\ee
(cf.~Eq.~(26) of~\cite{Beneke:2005vv}) has to be added to the 
diagrams in Figure~\ref{fig3} to obtain the final result 
for $H^{\rm II}_{ik}$, where $T^{\rm I(1)}_{ik}$ is the penguin 
contraction contribution to the 1-loop kernel $T^{I}_{ik}$. 
The form factor subtraction affects the structure of the result. 
For instance, we find that diagrams (e) and (f) in Figure~\ref{fig3} 
cancel exactly against the $C_{8g}$ term of the form factor 
subtraction, such that there is no dependence on $C_{8g}^{\rm eff}$ 
left, i.e. the primitive kernel $m_1=0$.

The ``tadpole'' contractions shown in Figure~\ref{fig3} (c) and (d) 
can be written as an effective flavour-changing two-point vertex 
\be
 \frac{G_F}{\sqrt{2}} \left[\,\sum_{p=u,c} \,V_{pD}^* V_{pb}\right]
(C_5+N_c C_6) \,A\,m_b^3\bar D (1+\gamma_5)b ,
\ee
where $A$ is a dimensionless, ultraviolet-divergent constant. 
However, the twelve tree diagrams with this vertex insertion 
vanish in their sum, so there is no contribution from
this class of diagrams. 
We also find that the diagram classes (b) and (f) 
in Figure~\ref{fig3} actually vanish. 

Penguin contractions to the hard-spectator scattering kernels relevant
to the QCD penguin amplitude $\alpha^p_4$ have 
previously been calculated in \cite{Li:2005wx}. In this paper 
the diagram classes (b), (c), (d) and (f) in Figure~\ref{fig3} 
are not considered (fortunately they all vanish), but diagrams 
(a5), (a6), (e3), (e4) and (e5) are also omitted, which is incorrect. The 
correct procedure includes these diagrams together with the 
form-factor subtraction discussed above. 
This seems to lead to significant numerical differences
between our results and those of \cite{Li:2005wx}.

\subsubsection*{Electroweak penguin contractions 
(Figure~\ref{fig1} (f) -- (m))}

Electroweak penguin contributions from topologies specified in the 
first two lines in Figure~\ref{fig3} can be obtained by
straightforward adjustment of colour factors from the corresponding
QCD penguin diagrams. 

The evaluation of the 
set of diagrams shown in the third line
of the figure, which contributes to $\alpha^p_{3,\rm EW}$, requires 
some explanation, since the photon has small virtuality. 
When the vector meson $M_2$ is transverse, this leads to an 
enhancement \cite{Beneke:2005we}, but for the longitudinal 
polarization state considered here the photon propagator 
cancels, and the diagrams contribute to the matching coefficients 
of SCET${}_{\rm I}$ four-quark operators as shown for the non-spectator 
kernels $T^{\rm I}$ in \cite{Beneke:2003zv}.

We calculate this contribution by keeping the up-going quark and 
anti-quark on-shell ($q_1^2=q_2^2=0$), but we assign them a 
small transverse momentum $q_{2\perp}=-q_{1\perp}$, 
leading to a non-zero virtuality of the intermediate photon. 
Then a straightforward calculation shows that we may perform 
the substitution
\be
\frac{-i}{q^2}\left(g^{\alpha\beta} - (1-\xi)\frac{q^\alpha
    q^\beta}{q^2}\right)\,\bar q(q_1)\gamma_\beta q(q_2) 
\to -2 i\left(\frac{q^\alpha}{m_b q^2}-\frac{n_-^\alpha}{m_b^2}
\right) \bar\chi\nmh\chi
\ee
in these diagrams. The $q^\alpha$ term contracts to zero with 
the remainder of the diagrams (after summing them all) as required by
gauge invariance; in the second term the photon pole is 
manifestly cancelled. Nevertheless, one cannot set $q^2=0$ at this 
point, since the loop integrations result in $\ln q^2$. These 
infrared logarithms are subtracted by performing a SCET matching 
calculation. With $q^2\not=0$ the SCET matrix elements do not 
vanish and contain $\ln (q^2/\nu^2)$. After matching 
the hard-scattering kernels 
are finite as $q^2\to 0$, but depend on the ultraviolet 
subtraction scale $\nu$, which is related to electromagnetic 
scale dependence of the longitudinal neutral vector meson decay constant. 
This scale dependence arises because (contrary to a widely held  
opinion) the electromagnetic current has an anomalous dimension 
in QED, which is precisely related to the penguin 
contractions as noted in 
\cite{Beneke:2003zv}. See also  \cite{Collins:2005nj} for an 
explanation of this point.

\subsection{Results for the primitive kernels}
\label{sec:primitive}

The analytical results for the one-loop kernels are as follows.

\subsubsection*{Vertex contractions (Figure~\ref{fig1} (a),(b))}

The kernels $r_1$ and $r_2$ can be found in~\cite{Beneke:2005vv}. 
The two new vertex kernels are related to these by
\bea
        r_3(u,v) &=& - r_1(\ub,v) + (4\, C_F - C_A) \Bigg( r_2(\ub,v)
        + \frac{3}{u} \Bigg) ,
\\
        r_4(u,v) &=& r_2(\ub,v) + \frac{6}{u}. 
\eea
The colour factors are $C_F=(N_c^2-1)/(2 N_c)=4/3$ and $C_A=N_c=3$.

\subsubsection*{QCD penguin contractions 
(Figure~\ref{fig1} (c) -- (e))}

All kernels can be expressed in terms of the four functions 
$s_1$, $s_3^F$, $s_3^A$, $s_4^\prime$. With 
\bea
        s_3^F(s) &=& \frac{14}{9} \frac{1}{\ub}  - \frac{1}{\ub v}
                     - \frac{8 (\ub + u v) s}{3 \ub^2 v}
                     - \frac{2 (1 - 2 v) s}{\ub v} C_0(s, \ub, v)
\nonumber \\ &&
                     + \Bigg( \frac{8 u s}{3 \ub^2} -
                                \frac{6 - 4 u \vb}{3 \ub \vb} \Bigg)
                           G_0(s, \ub)
                     + \Bigg( \frac{4}{3} + \frac{8 s}{3 \ub v} +
                                \frac{2}{\ub \vb} \Bigg) G_0(s, \ub v)
\nonumber \\ &&
                     + \frac{2}{3 \ub} \bigg(\ln s +
                     \ln\frac{m_b^2}{\mu^2} \bigg) ,
\label{s3F}\\
        s_3^A(s) &=& -\frac{5}{18} \frac{1}{\ub} + \frac{1}{3 \ub v}
                     + \frac{2 s}{\ub^2 v} + \frac{\vb s}{\ub v}
                                               C_0(s,\ub,v)
\nonumber \\ &&
                     + \Bigg( \frac{1}{3 \ub \vb}
                              - \frac{2 (3-v) s}{3 \ub^2 v \vb}
                          \Bigg) G_0(s, \ub)
                     + \Bigg( \frac{4 s}{3 \ub^2 \vb}
                              - \frac{1+2\vb}{3\ub\vb} \Bigg) G_0(s, \ub v)
\nonumber \\ &&
                     - \frac{1}{3\ub} \bigg(\ln s + 
                       \ln \frac{m_b^2}{\mu^2}\bigg),
\label{s3A}
\eea
we have
\bea
        s_1(s) &=& \frac{1}{\ub} - \frac{1}{2\,\ub  v}
                   - \frac{(1 - 2 v) s}{\ub v} C_0(s, \ub, v)
                   + \frac{1}{\ub \vb}  (G_0(s, \ub v) - G_0(s, \ub) ) ,
\\
        s_1'(s) &=& -s_1(s),
\\
        s_2(s) &=& s_3(s) - \Bigg(C_F - \frac{C_A}{2} \Bigg)
                       \frac{2}{3\ub} ,
\\
        s_2'(s) &=& s_3(s) - (4 C_F - C_A) s_1(s)
                       - \Bigg(C_F - \frac{C_A}{2} \Bigg)
                               \frac{2}{3\ub} ,
\\
        s_3(s) &=& C_F s_3^F(s) + C_A s_3^A(s) ,
\\
        s_3'(1) &=& \Bigg(2 C_F - \frac{C_A}{2} \Bigg) s_4'(1) ,
\\
        s_4(s) &=& s_1(s) ,
\\
        s_4'(1) &=& - \frac{2}{\ub v}
                    - \frac{2(2-\ub \bar v)}{\ub v} C_0(1, \ub, v)
                    - \frac{4}{\ub \vb} (G_0(1, \ub v) - G_0(1, \ub)) ,
\\
        m_1 &=& 0. 
\eea
Here
\bea
        &&G_0(s, x) = \frac{\sqrt{4 s - x - i \epsilon}}{\sqrt{x}}
                        \arctan\frac{\sqrt{x}}{\sqrt{4 s - x - i
                            \epsilon}} ,
\nonumber \\
        && C_0(s, x, y) = \frac{2}{x(1-y)} \Bigg(
             \arctan^2 \sqrt{\frac{x y}{4 s - x y - i \epsilon}}
             - \arctan^2 \sqrt{\frac{x}{4 s - x - i \epsilon}} \,\Bigg).
\eea

\subsubsection*{Electroweak penguin contractions 
(Figure~\ref{fig1} (f) -- (m))}

\bea
        s_{1,\rm EW}(s) &=& N_c \Bigg(s_3^F(s) - 2 s_1(s) -
                                \frac{2}{3\ub} \Bigg) ,
\\
        s'_{1,\rm EW}(s) &=& s_{1,\rm EW}(s) ,
\\
        s_{2,\rm EW}(s) &=& s_3^F(s) - \frac{2}{3\ub} ,
\\
        s'_{2,\rm EW}(s) &=& s_{2,\rm EW}(s) - 4 s_1(s) ,
\\
        s_{3,\rm EW}(s) &=& s_3^F(s) ,
\\
        s'_{3,\rm EW}(1) &=& 2 s_4'(1) ,
\\
        s_{4,\rm EW}(s) &=& N_c (s_3^F(s) - 2 s_1(s)) ,
\\
        s'_{4,\rm EW}(1) &=& 0 ,
\\
        m_{1,\rm EW} &=& 0 ,
\\
        m_{2,\rm EW} &=& 0 ,
\\
        s_{5,\rm EW}(s) &=& \left\{ \begin{array}{ll} 
            N_c \bigg(6 + 2 i \pi - 2 \ln \vb -
              2 \,{\displaystyle \ln \frac{m_b^2}{\nu^2}} \bigg) & 
              \quad (s = 0), \\[0.4cm]
            N_c \Big(6  - 4 G_0(s,\vb)+ 4 s C_0(s, \vb, 0) \Big) & 
              \quad (s \not= 0), \end{array} \right.
\\
        s_{8,\rm EW} (s) &=& s_{5,\rm EW}(s),
\\
        s'_{8,\rm EW}(s) &=& - s_{5, \rm EW}(s),
\\
        m_{3,\rm EW} &=& 4 N_c ,
\\
        m_{4,\rm EW} &=& 0. 
\eea
The kernels $s_{6,\rm EW}$, $s_{7,\rm EW}$, $s'_{7,\rm EW}$ vanish.
The kernels $s'_{5,\rm EW}$ and $s'_{6,\rm EW}$ vanish after the
rearrangement of Wilson coefficients discussed after (\ref{H3REW}). 
The scale $\nu$ in $s_{5,\rm EW}$ is due to the electromagnetic 
scale dependence of the neutral vector meson decay constant. In the 
following we no longer distinguish $\nu$ from the matching scale 
$\mu$ in the other expressions.

\section{Numerical penguin amplitudes}
\label{sec:pengamps}

In this section we study the numerical values of the calculated 
corrections to the penguin amplitudes $\alpha^p_{i}(M_1 M_2)$, 
$\alpha^p_{i,\rm EW}(M_1 M_2)$ ($i=3,4$), and provide updated 
results for some phenomenologically important 
penguin-to-tree and penguin-to-penguin amplitude ratios. 

\subsection{Input parameters}

\renewcommand{\arraystretch}{1.4}
\begin{table}
\begin{center}
\begin{tabular}{|cc|cc|}\hline
Parameter & Value/Range & Parameter & Value/Range \\
\hline
$\Lambda_{\overline{\mathrm{MS}}}^{(5)}$ & 0.225 & 
$\mu_b$ & $4.8^{+4.8}_{-2.4}$ \\
$m_c$ & $1.3\pm 0.2$  & 
$\mu_\mathrm{hc}$ & $1.5\pm 0.6$ \\
$m_s$(2 GeV) & $0.09\pm 0.02$ & 
$f_{B_d}$ & $0.21 \pm 0.02$  \\
$(m_u+m_d)/m_s$ & 0.0826 &
$f_{\pi}\,[f_K]$ & 0.131 [0.16]  \\
$m_b$ & 4.8 & 
$f^{B\pi}_+(0)$ &  $0.25 \pm 0.05$  \\
$\bar m_b(\bar m_b)$ & 4.2 & 
$f^{BK}_+(0)$ & $0.34\pm 0.05$ \\
$|V_{cb}|$ & $0.0415\pm 0.0010$  & 
$A_0^{B\rho}(0)$ & $0.32\pm 0.05$ \\
$|V_{ub}/V_{cb}|$ &  $0.09\pm 0.02$  & 
$\lambda_B$(1 GeV) & $0.35 \pm 0.15$  \\
$\gamma$ & $(70 \pm 20)^\circ$ & 
$\sigma_1$(1 GeV) &  $1.5 \pm 1$\\
$\tau(B^-)$ & 1.64\,\mbox{ps} &
$\sigma_2$(1 GeV) &  $3 \pm 2$\\
$\tau(B_d)$ & 1.53\,\mbox{ps} &
$a^{\bar K}_1$(2 GeV) &  $0.06 \pm 0.06$  \\
&&
$a^{\pi,\bar K}_2$(2 GeV) &  $0.2 \pm 0.15$  \\
\hline
\end{tabular}
\end{center}
\caption{List of input parameters. Dimensionful parameters are given
in units of 1 GeV. \label{tab:inputs}}
\end{table}

The calculation of non-leptonic decay amplitudes needs several input 
parameters, which we summarize in Table~\ref{tab:inputs}. These are 
fundamental parameters such as the strong coupling, quark masses 
and CKM parameters; the matching and renormalization scale parameters
$\mu_b$ and $\mu_{\rm hc}$ (see following subsection); and hadronic 
parameters related to decay constants, form factors and light-cone 
distribution amplitudes of the mesons. 

There has been significant progress in the calculation of the 
first few Gegenbauer moments of the pion and kaon light-cone distribution 
amplitudes from QCD sum rules and lattice 
QCD \cite{Khodjamirian:2004ga,Braun:2004vf}, which leads to a 
change in the corresponding entries in the table compared to earlier 
analyses.\footnote{Our kaon Gegenbauer moments are 
defined as $\langle K|\ldots|0\rangle$ matrix elements 
rather than  $\langle 0|\ldots|K \rangle$ and therefore the odd moments have 
opposite sign compared to those papers.} Less is known 
about the moments of the $B$ meson distribution amplitude, 
\begin{equation}
\frac{1}{\lambda_B(\mu)} \equiv \int_0^\infty
\frac{d\omega}{\omega}\,\phi_{B+}(\omega;\mu),
\qquad
\sigma_n(\mu) \equiv \lambda_B(\mu) \int_0^\infty \frac{d\omega}{\omega} 
\,\phi_{B+}(\omega;\mu)\,\ln^n \frac{\mu}{\omega}.
\label{lamb}
\end{equation}
The logarithmic moments enter only in the 1-loop correction 
to the hard-collinear kernel $J$ in (\ref{Xifact}). However, 
$\lambda_B$ is very important, because the 
hard spectator-scattering contribution to the decay amplitudes 
is directly proportional to $1/\lambda_B$. The value 
of $\lambda_B$ adopted 
here follows \cite{Beneke:1999br} and 
represents a compromise between QCD sum rules and models of 
the B meson distribution amplitude that seem to favour 
larger values \cite{Braun:2003wx} and data that favour 
smaller values \cite{Beneke:2005vv,Beneke:2003zv}. Recent 
evaluations of the $B\to\pi$ form factor at $q^2=0$ also 
tend to smaller values \cite{Ball:2004ye}, and we have 
therefore adjusted the input to the number taken in 
scenario S4 of \cite{Beneke:2003zv}. The difference in 
$F^{B\pi}(0)$ vs. $F^{BK}(0)$ is consistent with the large 
SU(3) breaking found in \cite{Khodjamirian:2004ga,Ball:2004ye}.

Parameters for mesons other than pions and kaons or parameters not given in 
the Table are taken from \cite{Beneke:2003zv}. 

\subsection{Renormalization group improvement and numerical 
implementation}

Our main task is to evaluate the integrals 
\be
I = -\frac{1}{2F^{B M_1}(0)}
\int_0^1 du dv \,H_{ik}^{{\rm II},p}(u,v;\mu_{\rm hc}) 
\,\hat\Xi_{M_1}(1-v;\mu_{\rm hc}) \phi_{M_2}(u;\mu_{\rm hc})
\label{int1}
\ee
in (\ref{eq:alphadef}) with $\hat\Xi_{M_1}(1-v;\mu_{\rm hc})$ given in 
(\ref{Xifact}). Hard spectator scattering depends on the 
two scales $\mu_b\sim m_b$ and 
$\mu_{\rm hc}\sim \sqrt{m_b \Lambda}$. In the previous 
equation we indicated the renormalization scale in 
the arguments of all quantities. To avoid formally large 
logarithms of the ratio of the two scales we need 
$H_{ik}^{{\rm II},p}$ at $\mu_{\rm hc}$ in this equation. 
On the other hand, our calculation refers to matching the 
effective weak Hamiltonian to SCET${}_{\rm I}$ at 
the scale $\mu_b$, i.e.\ the scale $\mu$ in Section~\ref{sec:primitive}
is $\mu_b$. In the following we explain the 
evolution of the hard-scattering coefficient to $\mu_{\rm hc}$, and 
the subsequent numerical evaluation of the integrals $I$. 

The renormalization-group evolution of 
the hard-scattering coefficients is the same for all coefficients 
and can be written as 
\begin{eqnarray} 
H_{ik}^{{\rm II},p}(u,v;\mu_{\rm hc}) &=& 
e^{-S(\mu_b,\mu_{\rm hc})} \int_0^1 du' dv'\, 
U_{\rm BL}(u,u';\mu_b,\mu_{\rm hc})
\nonumber\\
&&\hspace*{0cm} \times 
U_{\parallel}(1-v,1-v';\mu_b,\mu_{\rm hc}) 
H_{ik}^{{\rm II},p}(u',v';\mu_b).
\end{eqnarray}
This follows because the evolution function factorizes into a term from the 
Brodsky-Lepage kernel \cite{Lepage:1980fj} for the evolution of 
light-meson distribution amplitudes, and a Sudakov suppression 
factor times the evolution kernel for the so-called 
SCET B1-type currents \cite{Hill:2004if,Beneke:2005gs}. 
(The arguments of $U_\parallel$ follow the convention of 
\cite{Beneke:2005gs}, where they refer to the gluon momentum fraction 
in the B1-type operator,
which is $1-v$.) The $U_{\rm BL}$ factor can be used 
to evolve $\phi_{M_2}(u;\mu_{\rm hc})$ to the scale $\mu_b$, 
so (\ref{int1}) reads 
\begin{eqnarray}
I &=& -\frac{1}{2 F^{B M_1}(0)}\,e^{-S(\mu_b,\mu_{\rm hc})} \int_0^1 du dv' \, 
\phi_{M_2}(u;\mu_b)H_{ik}^{{\rm II},p}(u,v';\mu_b) 
\nonumber\\
&&\times 
\int_0^1 d v\,U_{\parallel}(1-v,1-v';\mu_b,\mu_{\rm hc})
\,\hat\Xi_{M_1}(1-v;\mu_{\rm hc}).
\label{int2}
\end{eqnarray}

The hard-scattering coefficients can be divided into 
two terms, 
\be 
H_{ik}^{{\rm II},p} = H_{Vik}^{{\rm II},p} + H_{Pik}^{{\rm II},p},
\ee 
referring to the vertex contractions (primitive kernels $r_k$) 
and penguin contractions (the others). The renormalization-group 
evolution and numerical evaluation of the first part is done exactly 
as described in Section 4.2 of \cite{Beneke:2005vv}. In the 
following we discuss only the second part. This part begins 
at order $\alpha_s$ (or $\alpha_{\rm em}$), hence it is 
sufficient to evaluate $\hat\Xi_{M_1}$ in the tree 
approximation (order $\alpha_s$) for the hard-collinear 
kernel $J$ resulting in 
\be
\hat\Xi_{M_1}(1-v;\mu_{\rm hc}) = -\frac{\pi C_F }{N_c}
\frac{\alpha_s(\mu_{\rm hc})f_{M_1}\hat f_B(\mu_{\rm hc})}
{m_b \lambda_B(\mu_{\rm hc})}
\frac{\phi_{M_1}(v;\mu_{\rm hc})}{\bar v}.
\ee 
Now we expand $\phi_{M_1}(v;\mu_{\rm hc})$ into Gegenbauer polynomials
\begin{equation}
\phi_{M_1}(v;\mu_{\rm hc})\,=\, 6 v\bar v 
\sum\limits_{n=0}^{\infty}
a_n^{M_1}(\mu_{\rm hc}) C^{(3/2)}_n(2 v-1),
\end{equation}
where $a_n^{M_1}(\mu_{\rm hc})$ are the Gegenbauer
moments ($a_0^{M_1}(\mu_{\rm hc})=1$), and define
\be
{\cal C}_n(v;\mu_b,\mu_{\rm hc}) = 
\int_0^1 dw\, 6 w C^{(3/2)}_n(2 w-1)\,
U_{\parallel}(1-w,1-v;\mu_b,\mu_{\rm hc}).
\label{defC}
\ee
Inserting this into (\ref{int2}) we obtain 
\begin{eqnarray}
I &=& 
\frac{f_{M_1}\hat f_B(\mu_{\rm hc})}
{m_b F^{B M_1}(0) \lambda_B(\mu_{\rm hc})}
\frac{\pi \alpha_s(\mu_{\rm hc}) C_F }{2 N_c}
\,e^{-S(\mu_b,\mu_{\rm hc})} 
\nonumber\\
&&\times 
\sum\limits_{n=0}^{\infty}a_n^{M_1}(\mu_{\rm hc})  \int_0^1 du dv \, 
\phi_{M_2}(u;\mu_b)\,H_{Pik}^{{\rm II},p}(u,v;\mu_b)\,
{\cal C}_n(v;\mu_b,\mu_{\rm hc}).
\label{int3}
\end{eqnarray}
It remains to calculate the ${\cal C}_n(v;\mu_b,\mu_{\rm hc})$. This
can be done by solving numerically the integro-differential 
equation 
\be
\mu \frac{d}{d\mu} \,{\cal C}_n(v;\mu,\mu_{\rm hc}) 
= - \int_0^1 dw \,\gamma_{\parallel}(1-v,1-w) \,
{\cal C}_n(w;\mu,\mu_{\rm hc})
\label{evolve}
\ee
subject to the initial condition 
${\cal C}_n(v;\mu_{\rm hc},\mu_{\rm hc}) = 6 v C^{(3/2)}_n(2 v-1)$, 
which follows from the defining equation (\ref{defC}) and the 
renormalization-group equation for $U_\parallel$. 
Here $\gamma_{\parallel}(\tau,\tau')$ is the leading-order 
anomalous dimension of the B1-type currents 
\cite{Hill:2004if,Beneke:2005gs} (Eq.~(99) in
\cite{Beneke:2005gs}). When this is done, the leading-logarithmic 
terms are summed to all orders, and (\ref{int3}) is free from 
formally large logarithms of the ratio of $m_b$ and 
$\sqrt{m_b\Lambda}$.

We truncate the Gegenbauer expansion of $\phi_{M_1}$ 
and $\phi_{M_2}$ after the second moment, so (\ref{int3}) is 
a sum of nine terms, each proportional to a product of two Gegenbauer
moments. For fixed $\mu_{\rm hc}$ and $\mu_b$ the evolution equation 
and the remaining integrals can be solved numerically. 
In practice, we wish to construct a code that allows us to 
vary the scales freely in order to estimate theoretical errors. 
The numerical evaluation is then too time-consuming. We therefore
proceed as follows. First we solve the evolution equation (\ref{evolve}) by 
successive approximation up to the second order. Since the
leading-order anomalous dimension $\gamma_\parallel$ depends 
on $\mu$ only through $\alpha_s(\mu)$, it is 
convenient to introduce a variable $x$ via  
\be 
\mu \frac{d}{d\mu} = \beta(\alpha_s) \frac{d}{d\alpha_s} =
- \frac{\alpha_s(\mu)}{4\pi} \frac{d}{dx}
\ee
with $x=0$ for $\mu=\mu_{\rm hc}$. Then we solve 
(\ref{evolve}) up to and including terms of order $x^2$. 
Given $\mu_b$ and $\mu_{\hc}$, the required value of $x$ is 
\be
x=\frac{1}{2\beta_0}\left(\ln\frac{\alpha_s(\mu_b)}
{\alpha_s(\mu_{\rm hc})}-\frac{\beta_1}{\beta_0}
\left[\frac{\alpha_s(\mu_b)}{4\pi}-\frac{\alpha_s(\mu_{\rm
      hc})}{4\pi}\right]\right)
\ee
with $\beta_{0,1}$ the first two coefficients of the beta-function, 
here for four flavours. It would be consistent to drop the 
$\beta_1$ terms, but since we always use 2-loop running of $\alpha_s$ in 
our code, we also include the corresponding 
terms here.  We have checked that the second-order approximation 
to the numerical solution of the evolution equation always 
provides an adequate approximation, even for the largest 
scale hierarchy allowed by Table~\ref{tab:inputs}, when 
$x$ reaches about $-0.06$. 

We thus obtain ${\cal C}_{0,1,2}(v;x)$ as second-order polynomials 
in $x$, which can be integrated numerically coefficient by coefficient in 
(\ref{int3}). A final complication arises due to the dependence of 
some of the $H_{Pik}^{{\rm II},p}(u,v;\mu_b)$ on the charm 
quark mass, more precisely on $m_c/m_b$, which should also 
be allowed to be variable. We deal with this by integrating (\ref{int3}) 
for several values of $m_c/m_b$, and then generate a polynomial fit 
which approximates the dependence on $m_c/m_b$ in the relevant 
interval. Thus the final result for $I$ is represented as 
a sum of nine terms, corresponding to the coefficients of 
products of Gegenbauer moments. Each term is a second-order polynomial 
in $x$, and in some cases also a polynomial in $m_c/m_b$ 
with numerical coefficients, such that the dependences on 
all parameters can be rapidly evaluated. 

The numerical results below include the renormalization-group 
evolution as described here. We have, however, found that the 
effect of summing logarithms is not very important. A simpler 
implementation leaving out evolution would not lead to 
essential differences in the sense that any difference to the 
evolved results is smaller than other uncertainties.  

\subsection{Amplitudes}

The final result for the $\alpha_i^p$ parameters in 
(\ref{eq:alphadef}), now including all contributions, 
can be written in a notation similar to 
Eq.~(35) of~\cite{Beneke:2003zv}, which makes explicit the contributions 
from the form-factor term in the factorization formula (\ref{eq:qcdfact}), 
from hard spectator-scattering, separately from tree- and 1-loop corrections, 
and from penguin contractions. To simplify the notation we 
use the identification $a_1=\alpha_1$, $a_2=\alpha_2$, 
$a^p_3=\alpha^p_{3L}$, $a^p_4=\alpha^p_{4L}$, $a^p_5=\alpha^p_{3R}$, 
$a^p_7=\alpha^p_{3R,\rm EW}$, $a^p_9=\alpha^p_{3L,\rm EW}$, 
$a^p_{10}=\alpha^p_{4L,\rm EW}$. We also give the coefficients 
$a^p_{6,8}$ related to the power-suppressed ``scalar'' penguin 
amplitudes for completeness, even though the 1-loop 
spectator-scattering correction has not yet been 
computed, and add the tree-level twist-3 spectator scattering term
to conform with the conventions of~\cite{Beneke:2003zv}. The general formula is 
\bea 
a_i^p(M_1 M_2) &=& C_i+\frac{C_{i\pm 1}}{N_c} + 
\frac{C_{i\pm 1}}{N_c}\,\frac{\alpha_s C_F}{4\pi}\,V_i(M_2) 
+P_i^p(M_2)
\nonumber\\
&&\hspace*{-1cm} +\,\frac{\pi\alpha_s C_F}{N_c^2}\,\frac{9 f_{M_1}\hat f_B}
{m_b F^{B M_1}(0) \lambda_B}\,
\bigg[C_{i\pm 1} \,H_i(M_1 M_2) 
+ \frac{\alpha_s}{4\pi}\bigg(
C_{i\pm 1} \,HV_i^{(1)}(M_1 M_2) 
\nonumber\\
&& \hspace*{0cm}
+ \,C_i \,HV_i^{(2)}(M_1 M_2)\bigg) 
+H\!P_i^p(M_1 M_2)\bigg].\qquad\qquad
\label{apar}
\eea
(Recall that $F^{B M_1}(0)=f_+^{BM_1}(0)$ for pseudoscalar $M_1$ and 
$A_0^{BM_1}(0)$ for $M_1=V$.) 
The upper signs apply when $i$ is odd, 
the lower ones when $i$ is even. The 
non-spectator terms $V_i(M_2)$, $P_i^p(M_2)$ are listed 
in \cite{Beneke:2003zv} (see also \cite{Beneke:2001ev}). 
The tree-level spectator-scattering 
term $H_i(M_1 M_2)$ differs from the convention of 
\cite{Beneke:2003zv} by an overall factor. Introducing 
\begin{equation}
\Delta_M \equiv \int_0^1 dx \,\frac{\phi_M(x)}{3 x} 
= 1 +\sum_{n=1}^\infty (-1)^n\,a_n^M,
\quad 
\bar\Delta_M \equiv \int_0^1 dx \,\frac{\phi_M(x)}{3 \bar x} 
= 1 +\sum_{n=1}^\infty a_n^M,
\end{equation}
we have
\begin{equation}
H_i(M_1 M_2) = \left\{
\begin{array}{ll} 
\displaystyle \phantom{-} \bar \Delta_{M_1}\bar \Delta_{M_2}+
\frac{1}{3} \,r_\chi^{M_1}\Delta_{M_2} X_H & 
 \qquad i=1-4,9,10\\[0.3cm]
\displaystyle - \bigg[\bar \Delta_{M_1} \Delta_{M_2} + 
\frac{1}{3} \,r_\chi^{M_1}\bar \Delta_{M_2} X_H\bigg] 
 & \qquad i=5,7\\[0.3cm]
\phantom{-} 0 & \qquad i=6,8.
\end{array}\right.
\label{sptree}
\end{equation}
Here $X_H$ models a non-factorizable power correction (since 
$r_\chi^{M_1} \sim 1/m_b$) as defined in \cite{Beneke:2003zv} 
[Eqs. (50) and (63)]. The new 1-loop spectator-scattering correction 
is contained in the objects $HV_i^{(1)}(M_1 M_2)$, 
$HV_i^{(2)}(M_1 M_2)$, $H\!P_i(M_1 M_2)$, which we now summarize 
in terms of the previously defined $H_{ik}^{\rm II}$ 
[Eq.~(\ref{H4L}--\ref{H3REW})] and primitive kernels. The vertex 
contractions (primitive kernels $r_i$) contribute 
\begin{eqnarray}
HV_i^{(1)}(M_1 M_2) &=& \left\{
\begin{array}{cl}
R_1(M_1 M_2)+\bar \Delta_{M_2} J(M_1) & 
 \qquad i=1-4,9,10\\[0.1cm]
R_3(M_1 M_2) - \Delta_{M_2} J(M_1)& \qquad i=5,7\\[0.1cm]
\mbox{unknown} & \qquad i=6,8,
\end{array}\right.
\\[0.3cm]
HV_i^{(2)}(M_1 M_2) &=& \left\{
\begin{array}{cl}
R_2(M_1 M_2) & 
 \qquad i=1-4,9,10\\[0.1cm]
R_4(M_1 M_2) & \qquad i=5,7\\[0.1cm]
\mbox{unknown} & \qquad i=6,8.
\end{array}\right.
\end{eqnarray}
Here $R_i(M_1 M_2)$ denotes the integrated vertex-contraction 
kernels 
\be
R_i(M_1 M_2) \equiv \frac{1}{9} \int_0^1 du dv\,
\phi_{M_1}(v)\, \phi_{M_2}(u) \,\frac{r_i(u,v)}{\bar v}.
\ee
and the next-to-leading order hard-collinear correction, $J(M_1)$, 
is given in Eq.~(50) of~\cite{Beneke:2005vv}. The above
expression for $R_i(M_1 M_2)$ does not include the renormalization-group 
summation of logarithms, since the final formula is
complicated. The contributions from the penguin contractions to 
spectator scattering are simply (\ref{int3}) up to a normalization 
factor, 
\begin{eqnarray}
H\!P_i^p(M_1 M_2)&=& 
e^{-S(\mu_b,\mu_{\rm hc})} \,\frac{1}{9}
 \sum\limits_{n=0}^{\infty}a_n^{M_1}(\mu_{\rm hc})  
\nonumber\\
&&\times\int_0^1 du dv \, 
\phi_{M_2}(u;\mu_b)\,\left[\frac{N_c}{2}\,H_{Pik}^{{\rm
      II},p}(u,v;\mu_b)
\right]\,
{\cal C}_n(v;\mu_b,\mu_{\rm hc}),
\label{int4}
\end{eqnarray}
which includes the log-summation. Again, this result does not 
apply to the power-suppressed penguin amplitudes, $i=6,8$, for 
which the radiative corrections are currently not known. 

\subsubsection{\boldmath The tree amplitudes $\alpha_{1,2}$}

The ``tree'' amplitudes $a_{1,2}(\pi\pi)$ have already been 
discussed in~\cite{Beneke:2005vv}. With our up-dated input 
parameters, their values read
\bea
a_1(\pi\pi) &=& 1.015 + [0.025 + 0.012i]_V 
   \nonumber\\ 
   && -\,\left[\frac{r_{\rm sp}}{0.485} \right]
   \Big\{ [0.020]_{\rm LO} + [0.034 + 0.029i]_{HV} + [0.012]_{\rm tw3} \Big\}
   \nonumber \\
   &=& 0.975^{+0.034}_{-0.072} + (-0.017^{+0.022}_{-0.051})i,
\\[0.2cm]
a_2(\pi\pi) &=& 0.184 - [0.153 + 0.077i]_V 
   \nonumber\\
   &&+ \,\left[ \frac{r_{\rm sp}}{0.485} \right]
   \Big\{ [0.122]_{\rm LO} + [0.050 +0.053i]_{HV} + [0.071]_{\rm tw3} \Big\}
   \nonumber \\
   &=& 0.275^{+0.228}_{-0.135} + (-0.024^{+0.115}_{-0.081})i.
\eea
In these expressions we separated the tree ($\alpha_s^0$, first number), 
vertex correction ($\alpha_s$, indexed by $V$) and 
the spectator-scattering correction 
(remainder). The latter is further divided into the tree ($\alpha_s$, 
indexed LO), 1-loop ($\alpha_s^2$, indexed $HV$), and 
twist-3 power correction (the $X_H$ term in (\ref{sptree})). 
The theoretical error in the last line of each expression is  
computed from the ranges in Table~\ref{tab:inputs} added 
in quadrature together with the error from $X_H$, which is treated 
as described in~\cite{Beneke:2003zv}. The most important 
parameter uncertainty is encoded in the combination 
\be
r_{\rm sp} = \frac{9 f_{M_1}\hat f_B}
{m_b F^{B M_1}(0) \lambda_B},
\ee
which normalizes the spectator-scattering term as can be seen 
from (\ref{apar}). 

The dependence on the final state mesons is rather small. For instance, 
the difference between $a_{1,2}(\pi K)$ and 
$a_{1,2}(\pi\pi)$, often called `non-factorizable' SU(3) breaking, 
is about 10\% for $a_2$ and the imaginary part of $a_1$ 
(and much smaller for the real part of $a_1$), 
since the recent estimates of the first Gegenbauer
moment of the kaon light-cone distribution amplitude favour 
small values. The dominant SU(3)-breaking effect can thus be 
estimated from the above expressions by the dependence of 
$r_{\rm sp}$ on $f_{M_1}/F^{BM_1}(0)$.

\subsubsection{\boldmath The QCD penguin amplitude $\alpha_4^p$}
\label{sec:alpha4p}

The QCD penguin amplitudes $a_4^{u,c}$ receive a 1-loop spectator-scattering
correction due to the primitive kernels $r_1,r_2$ and the newly
computed $s_1 \dots s_4$. We find
\bea
   a_4^u(\pi \pi) &=& -0.029 - [0.002 + 0.001i]_V + [0.003-0.013i]_P
\nonumber \\
&&   +\left[ \frac{r_{\rm sp}}{0.485} \right]
   \Big\{ [0.001]_{\rm LO} + [0.001 +0.001i]_{HV} - [0.000+0.001i]_{HP}
   + [0.001]_{\rm tw3} \Big\}
\nonumber \\
&=& -0.024^{+0.004}_{-0.002} + (-0.012^{+0.003}_{-0.002})i
\\[0.2cm]
   a_4^c(\pi \pi) &=& -0.029 - [0.002 + 0.001i]_V - [0.001+0.007i]_P
\nonumber \\
&&   +\left[ \frac{r_{\rm sp}}{0.485} \right]
   \Big\{ [0.001]_{\rm LO} + [0.001 +0.001i]_{HV} + [0.000-0.000i]_{HP}
   + [0.001]_{\rm tw3} \Big\}
\nonumber \\
&=& -0.028^{+0.005}_{-0.003} + (-0.006^{+0.003}_{-0.002})i
\eea
Here ``$P$'' and ``$HP$'' denote the contributions from 
penguin-contraction diagrams to the form-factor and spectator-scattering 
term in the factorization formula.
Numerically the new corrections, labeled ``$HV$'' and ``$HP$'',
respectively, are very small.\footnote{The terms labelled `$P$' and 
`$HP$' contain the charm penguin contractions. There is no evidence 
from this calculation that these effects are large.}  
As the large Wilson coefficient $C_1$ is
involved in the new penguin correction (the term $C_1 s_3(s_p)$ in 
(\ref{H4L})), this result is somewhat
surprising. Closer inspection shows that
\bea
   [a_4^u(\pi \pi)]_{HP} &=&
        [0.0074 + 0.0060i]_{C_A C_1} - [0.0073+0.0053i]_{C_F C_1}
        - 0.0002+0.0001 i,
\nonumber \\[0.2cm]  
   [a^{c}_4(\pi \pi)]_{HP} &=& 
        [0.0039-0.0000i]_{C_A C_1} - [0.0034-0.0006i]_{C_F C_1}
        - 0.0004-0.0002 i,
\nonumber
\eea
where the terms labeled ``$C_A C_1$'' and ``$C_F C_1$'' 
arise from the term  $C_1 s_3(s_p)$ separated according to the two colour
structures $C_A$ and $C_F$ (primitive kernels $s_3^A(s_p)$ and $s_3^F(s_p)$). 
Thus, there is an almost complete cancellation between the two 
terms, which individually would have resulted in a large correction 
to the QCD penguin amplitude. Inspecting (\ref{s3F}), (\ref{s3A}) 
some terms appear with the small colour factor $C_F-C_A/2$, but 
it is unclear to us whether there is an explanation for the near-completeness 
of the  numerical cancellation or whether it is accidental. 

\begin{figure}[t]
\centerline{
\includegraphics[width=11cm]{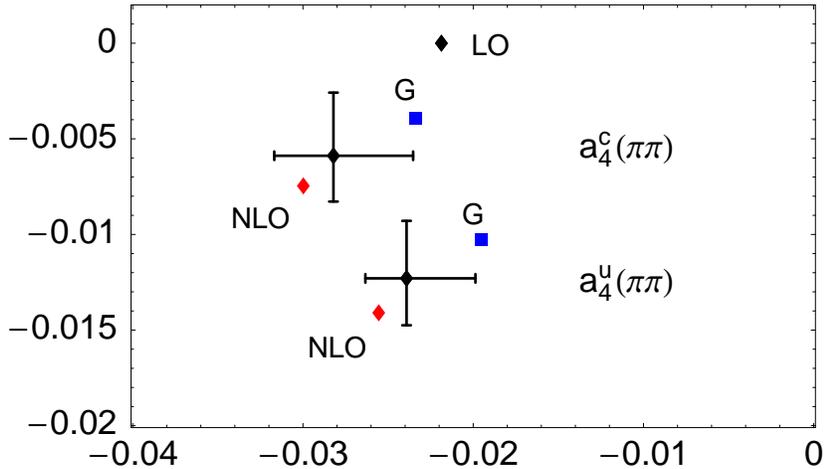}
}
\caption{\label{a4plot} LO, NLO and NLO${}_{
\rm sp}$ value of $a_4^p$ in the complex plane. 
$a_4^u$ has a sizeable phase. The NLO${}_{
\rm sp}$ point includes a theoretical error estimate.
}
\end{figure}

A graphical representation of 1-loop corrections to $a^p_4$
is given in Figure~\ref{a4plot}. The leading-order (LO) point 
corresponds to naive factorization. The point labelled ``NLO'' 
includes all $\alpha_s$ corrections. The point including the 
error bars is the result of our calculation that adds the 
1-loop spectator-scattering correction. We refer to this 
partial next-to-next-to-leading order (NNLO) calculation as 
``NLO${}_{\rm sp}$'', since the NNLO term is a 1-loop correction 
in the spectator-scattering sector. 
The point ``G'' corresponds to a choice of inputs
made in~\cite{Beneke:2005vv} to achieve an improved agreement with
the $B\to \pi \pi$ decay rates. This input parameter set uses 
$\lambda_B=200\,\mbox{MeV}$ and $a_2^\pi=0.3$ near the boundaries 
of the assumed regions instead of the central values given 
in Table~\ref{tab:inputs}, which leads to an increase of the 
spectator-scattering contribution.

The figure shows that the new correction has a small effect 
on the real part, but can be relevant to the imaginary part 
depending on parameter values. As was
found in~\cite{Beneke:2005vv} for the parameters $a_1,a_2$,
perturbation theory is well-behaved. These observations apply 
to all final states, in particular, ``non-factorizable'' SU(3) 
breaking turns out to be quite small, as discussed above. 

To gauge the impact of the spectator-scattering correction on 
branching fractions and CP asymmetries, we recall that $a_4^p$ 
always appears in conjunction with the power-suppressed 
penguin amplitude $r_\chi a_6^p$. Since we have not computed 1-loop 
corrections to the short-distance coefficients of power-suppressed 
operators, the value of  $r_\chi a_6^p$ differs from previous 
analyses only due to the updated input parameters. 
Furthermore, since the tree-level spectator scattering contribution
vanishes for $a_6^p$, the following numbers are solely due to the
naive-factorization and 1-loop contributions to the form-factor 
term in the factorization formula. We have
\bea
   r_\chi^\pi a_6^u(\pi \pi) &=& -0.055 - [0.001]_V - [0.004+0.020i]_P
\nonumber \\
&=&  -0.060^{+0.011}_{-0.017} + (-0.020^{+0.005}_{-0.006})i , \qquad
\\
   r_\chi^\pi a_6^c(\pi \pi) &=& -0.055 - [0.001]_V - [0.009+0.010i]_P
\nonumber \\
&=&
 -0.065^{+0.012}_{-0.019} + (-0.010^{+0.004}_{-0.004})i . \qquad
\eea
It is worth noting that $r_\chi^\pi a_6^p(\pi \pi)$ is numerically 
larger than $a^p_4(\pi\pi)$. On the other hand,  $r_\chi^\pi a_6^p(M_1 M_2)$ 
is strongly suppressed when $M_2$ is a vector meson, where the 
naive-factorization contribution is zero. 
We shall discuss some of the phenomenology associated 
with the penguin amplitudes in Section~\ref{sec:ratios}.

\subsubsection{\boldmath The flavour-singlet 
QCD penguin amplitude $\alpha_3^p$}

The flavour-singlet QCD penguin amplitude is not relevant to the 
$\pi\pi$ final state. We therefore give numerical values for 
\bea
   a_3^p(\bar K \phi) &=& 0.001 + [0.005 + 0.002i]_V
\nonumber \\
&&   -\left[ \frac{r_{\rm sp}}{0.435} \right]
    \Big\{ [0.003]_{\rm LO} + [0.001 +0.001i]_{HV} + [0.002]_{\rm tw3}
   \Big\} 
\nonumber \\
&=& 0.001^{+0.004}_{-0.005} + (0.001^{+0.002}_{-0.003}) i ,
\label{a3p}\\
   a_5^p(\bar K \phi) &=& -0.005 - [0.002 + 0.003i]_V
\nonumber \\
&&   +\left[ \frac{r_{\rm sp}}{0.435} \right]
    \Big\{ [0.004]_{\rm LO} + [0.002 +0.002i]_{HV} + [0.002]_{\rm tw3} \Big\}
\nonumber \\
&=& 0.001^{+0.007}_{-0.004} + (-0.000^{+0.005}_{-0.003}) i, 
\label{a5p}
\eea
valid for $p=u$ and $p=c$. Here the 1-loop spectator scattering 
term ``$HV$'' is numerically of the same size as the other 
terms, and gives a large effect. This is typical for corrections 
to the smaller colour-suppressed amplitudes. 
We recall from Section~\ref{sec:setup} that the calculation of 
the flavour-singlet amplitudes $a^p_{3,5}$ is not complete when 
$M_2$ is a pseudoscalar meson, since 
we did not compute the matching coefficients of two-gluon 
operators. Even more important conceptually, there is another 
term in the QCD factorization formula 
which does not factorize in the form of (\ref{eq:qcdfact}), which 
has been estimated to be of similar size as the 
individual terms in (\ref{a3p}), 
(\ref{a5p}) \cite{Beneke:2002jn}, but which is very uncertain. 

Because the flavour-singlet amplitude $\alpha_3^p=a_3\mp a_5$ is so 
small, it does not play an important role in the phenomenology of 
the more prominent charmless final states, in particular in the 
explanation of the large $\eta' K$ branching fractions \cite{Beneke:2002jn}. 
The large spectator-scattering correction found above affects final states 
such as $\eta' K^*$ or such with very small branching fractions, 
where the leading contribution from $\alpha_4^p$ 
is suppressed. However, the predictions for such decays carry 
large theoretical uncertainties.

\subsubsection{Electroweak penguin amplitudes}

The electroweak (EW) penguin amplitudes are small, yet they contribute 
significantly to isospin-breaking effects in decays to $\pi K$ final 
states. The ``colour-allowed'' EW penguin amplitude is 
$\alpha_{3,\rm EW}^p = a_9^p \mp a_7^p$ (upper sign for $\pi K$). 
We find 
\bea
a_7^u(\bar K \pi) &=&  10^{-2} \times
   \Bigg( 0.009 + [0.002 + 0.003i]_V + [0.038]_P
\nonumber \\
&&  \hspace*{-1cm} 
- \left[ \frac{r_{\rm sp}}{0.435} \right]
   \Big\{ [0.005]_{\rm LO} + [0.003 +0.004i]_{HV} - [0.020+0.016i]_{HP}
   + [0.003]_{\rm tw3} \Big\} \Bigg)
\nonumber \\
&& \hspace*{-1cm} = 10^{-2} \times
      \Big( 0.058^{+0.024}_{-0.017} + (0.015^{+0.010}_{-0.006})i \Big),
\\[0.2cm]
   a_7^c(\bar K \pi) &=&  10^{-2} \times
   \Bigg( 0.009 + [0.002 + 0.003i]_V + [0.011]_P
\nonumber \\
&& \hspace*{-1cm}  - \left[ \frac{r_{\rm sp}}{0.435} \right]
   \Big\{ [0.005]_{\rm LO} + [0.003 +0.004i]_{HV} + [0.002-0.001i]_{HP}
   + [0.003]_{\rm tw3} \Big\} \Bigg)
\nonumber \\
&& \hspace*{-1cm} = 10^{-2} \times
      \Big( 0.010^{+0.011}_{-0.017} + (0.000^{+0.003}_{-0.006})i \Big),
\\[0.2cm]
   a_9^u(\bar K \pi) &=&  10^{-2} \times
   \Bigg( -0.909 - [0.023 + 0.011i]_V + [0.038]_P
\nonumber \\
&& \hspace*{-1cm}  + \left[ \frac{r_{\rm sp}}{0.435} \right]
   \Big\{ [0.017]_{\rm LO} + [0.028 +0.025i]_{HV} + [0.020+0.016i]_{HP}
   + [0.010]_{\rm tw3} \Big\} \Bigg)
\nonumber \\
&& \hspace*{-1cm} = 10^{-2} \times
      \Big( -0.819^{+0.080}_{-0.042} + (0.029^{+0.053}_{-0.023})i \Big),
\\[0.2cm]
   a_9^c(\bar K \pi) &=&  10^{-2} \times
   \Bigg( -0.909 - [0.023 + 0.011i]_V + [0.011]_P
\nonumber \\
&& \hspace*{-1cm}  + \left[ \frac{r_{\rm sp}}{0.435} \right]
   \Big\{ [0.017]_{\rm LO} + [0.028 +0.025i]_{HV} - [0.002-0.001i]_{HP}
   + [0.010]_{\rm tw3} \Big\} \Bigg)
\nonumber \\
&& \hspace*{-1cm} = 10^{-2} \times
      \Big( -0.868^{+0.058}_{-0.026} + (0.015^{+0.043}_{-0.018})i \Big).
\eea
The penguin-contraction term ``$HP$'' originates from the diagrams shown 
in the third line of Figure~\ref{fig3}. Both ``$P$'' and  ``$HP$'' 
coincide for $a_7^p$ and $a_9^p$, such that they exactly cancel in 
the combination $a_9-a_7$ appropriate to final states with pseudoscalar 
$M_2$, but add up for vector $M_2$. The 1-loop spectator scattering 
correction is quite important for $a_7^p$, and $a_7^u$ is seen to 
even be dominated by the penguin contractions. However, what is
relevant to decay amplitudes is
$\alpha_{3,\rm EW}$, and therein
the 1-loop corrections are overwhelmed by the large tree contribution 
to $a_9^p$. 

\begin{figure}[t]
\centerline{\hskip-0.5cm
\includegraphics[width=11cm]{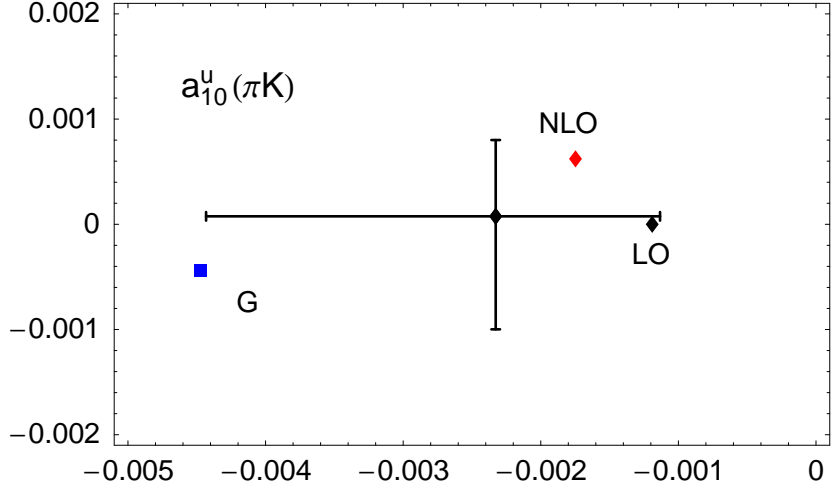}
}
\caption{\label{a10fig} The electroweak penguin amplitude $a^p_{10}$
receives a large correction. The amplitude is almost identical
for $p=u$ and $p=c$, therefore only the case $p=u$ is shown.}
\end{figure}

The colour-suppressed EW penguin amplitude $\alpha_{4,\rm EW}^p=
a_{10}^p \pm r_\chi a_8^p$ consists of $a_{10}^p$ and a power-suppressed 
amplitude $r_\chi a_8^p$, which is the electroweak equivalent to 
$r_\chi a_6^p$. We find 
\bea
   r_\chi^{\bar K} a_8^u(\pi \bar K) &=&  10^{-2} \times
   \Bigg( 0.063 + [0.000]_V + [0.004 - 0.016i]_P \Bigg)
\nonumber \\
&=& 10^{-2} \times
      \Big( 0.068^{+0.025}_{-0.024} + (-0.016^{+0.010}_{-0.012})i \Big),
\\
   r_\chi^{\bar K} a_8^c(\pi \bar K) &=&  10^{-2} \times
   \Bigg( 0.063 + [0.000]_V + [0.002 - 0.010i]_P \Bigg)
\nonumber \\
&=& 10^{-2} \times
      \Big( 0.065^{+0.025}_{-0.025} + (-0.010^{+0.006}_{-0.007})i \Big),
\\
   a_{10}^u(\pi \bar K) &=&  10^{-2} \times
   \Bigg( -0.161 + [0.135 + 0.073i]_V + [0.026 - 0.011i]_P
\nonumber \\
&& \hspace*{-1cm}  - \left[ \frac{r_{\rm sp}}{0.485} \right]
   \Big\{ [0.115]_{\rm LO} + [0.050 + 0.051i]_{HV} + [0.007 + 0.004i]_{HP}
   + [0.061]_{\rm tw3} \Big\} \Bigg)
\nonumber \\
&& \hspace*{-1cm} = 10^{-2} \times
      \Big( -0.233^{+0.119}_{-0.210} + (0.008^{+0.073}_{-0.107})i \Big),
\\
   a_{10}^c(\pi \bar K) &=&  10^{-2} \times
   \Bigg( -0.161 + [0.135 + 0.073i]_V + [0.023 - 0.006i]_P
\nonumber \\
&&   \hspace*{-1cm} - \left[ \frac{r_{\rm sp}}{0.485} \right]
   \Big\{ [0.115]_{\rm LO} + [0.050 + 0.051i]_{HV} + [0.002 - 0.001i]_{HP}
   + [0.061]_{\rm tw3} \Big\} \Bigg)
\nonumber \\
&& \hspace*{-1cm} = 10^{-2} \times
      \Big( -0.231^{+0.118}_{-0.207} + (0.017^{+0.071}_{-0.105})i \Big).
\eea
The coefficient $a_{10}^p$ is similar to the colour-suppressed tree 
amplitude $a_2$. The non-spectator term (first line in each expression) 
is strongly suppressed due to a cancellation between the naive-factorization 
term and the 1-loop correction, which is large due to the absence 
of colour-suppression. The final number is largely from 
spectator-scattering which obtains significant 1-loop corrections. 
The result is displayed graphically in Figure~\ref{a10fig}.

\section{Amplitude ratios}
\label{sec:ratios}

With the improved $a_i^p$ parameters we are in a position to calculate 
complex ratios of strong amplitudes such as $P/T$ or $C/T$. In this 
section we discuss a few prominent examples.

\subsection{\boldmath The $PP$, $PV$, $V\!P$, and $[VV]_L$ QCD penguin amplitude}
\label{sec:PPPV}

\begin{figure}[p]
    \vspace{0cm}
\centerline{
\includegraphics[width=7.4cm]{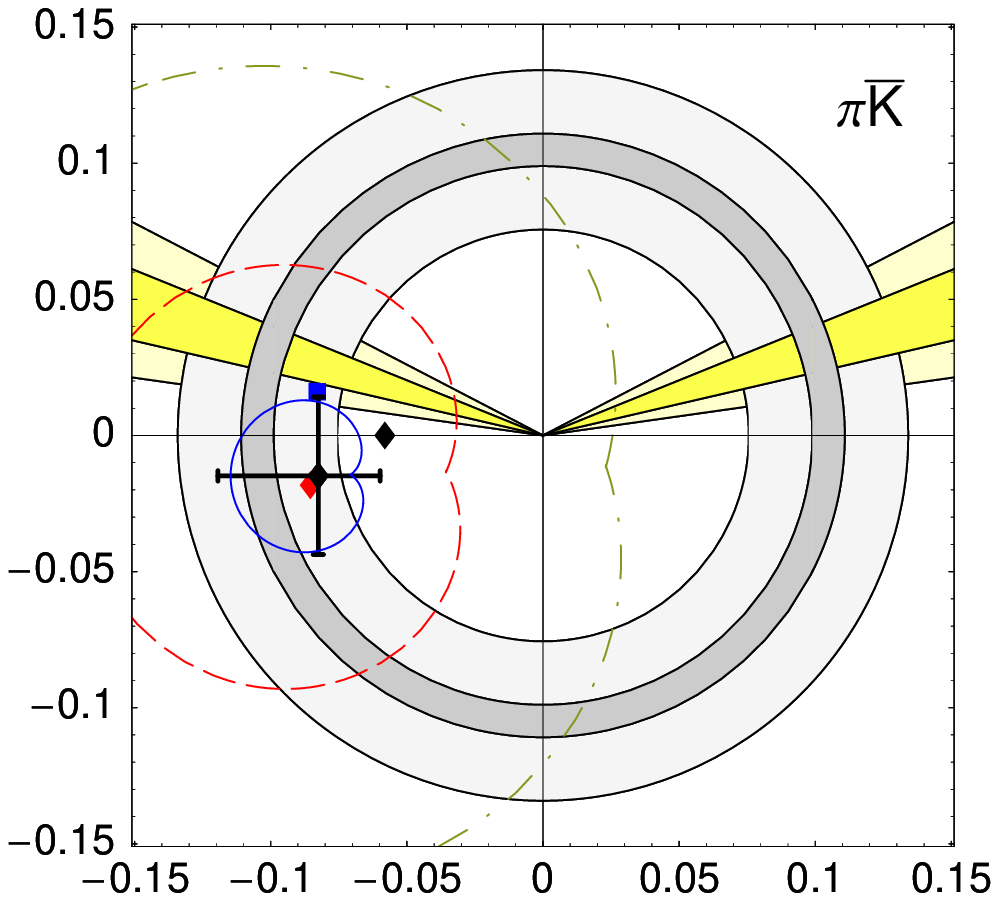}
\hskip5mm
\includegraphics[width=7.4cm]{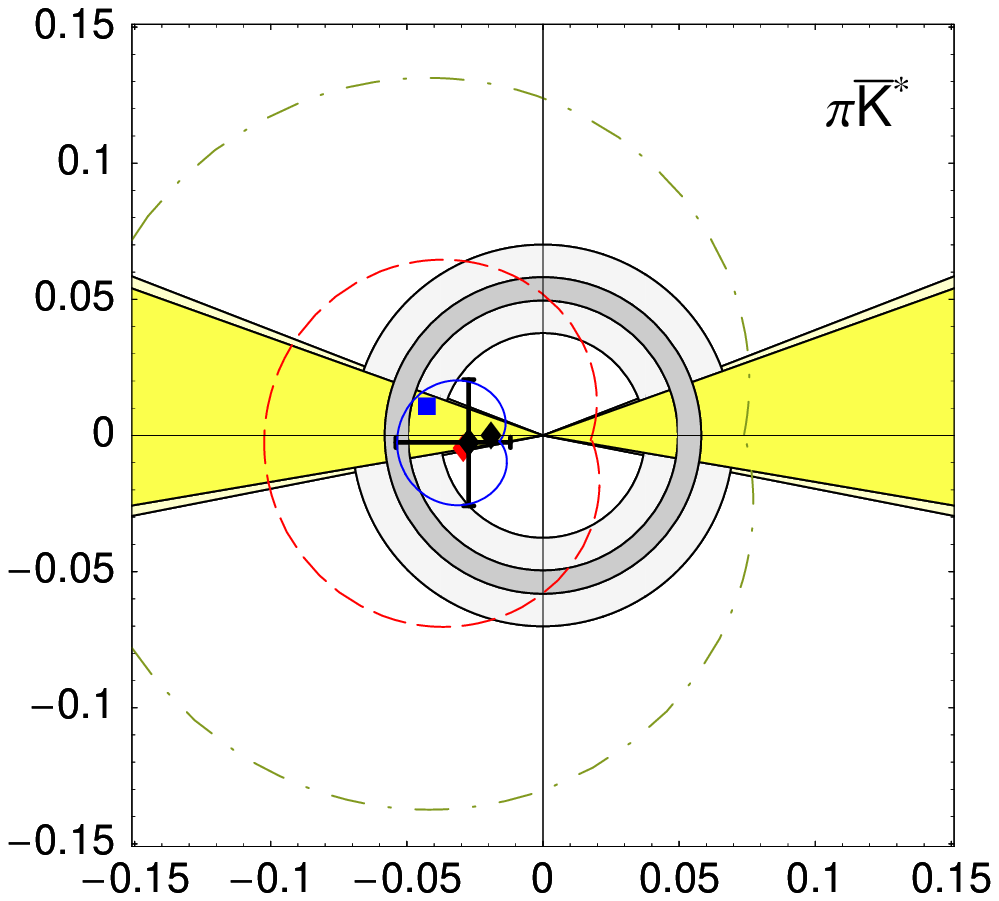}
}
\centerline{
\includegraphics[width=7.4cm]{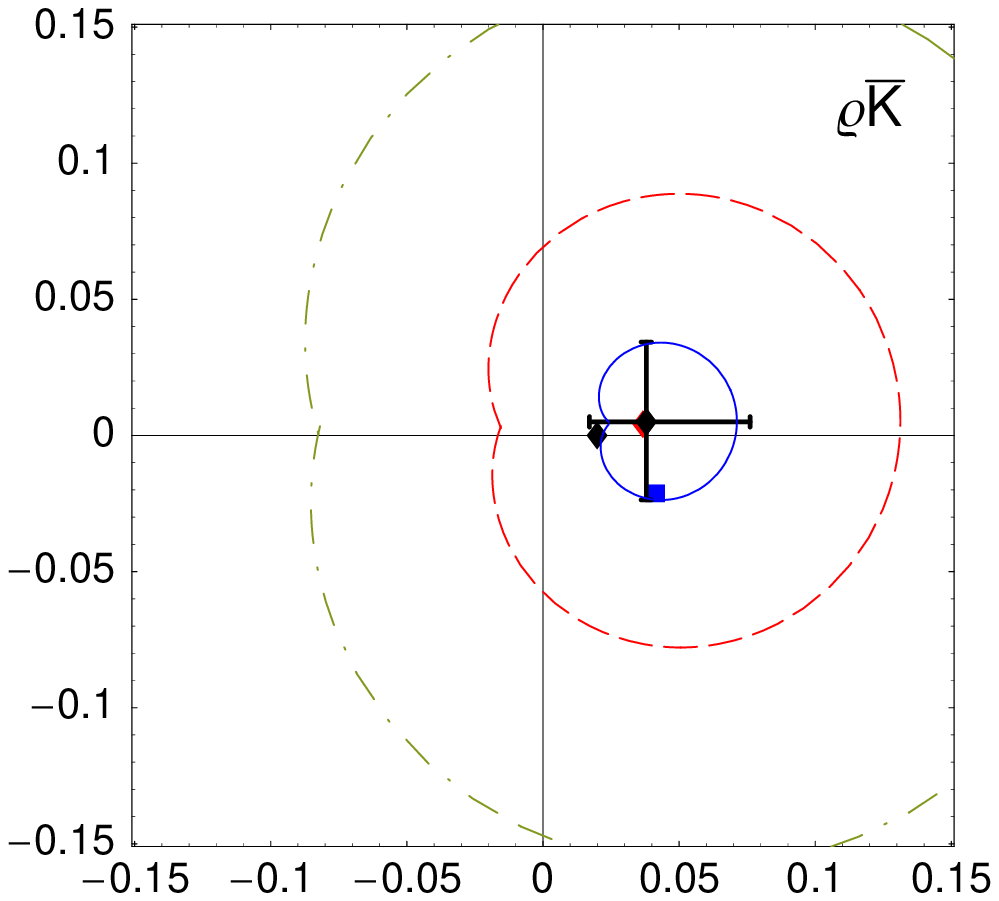}
\hskip5mm
\includegraphics[width=7.4cm]{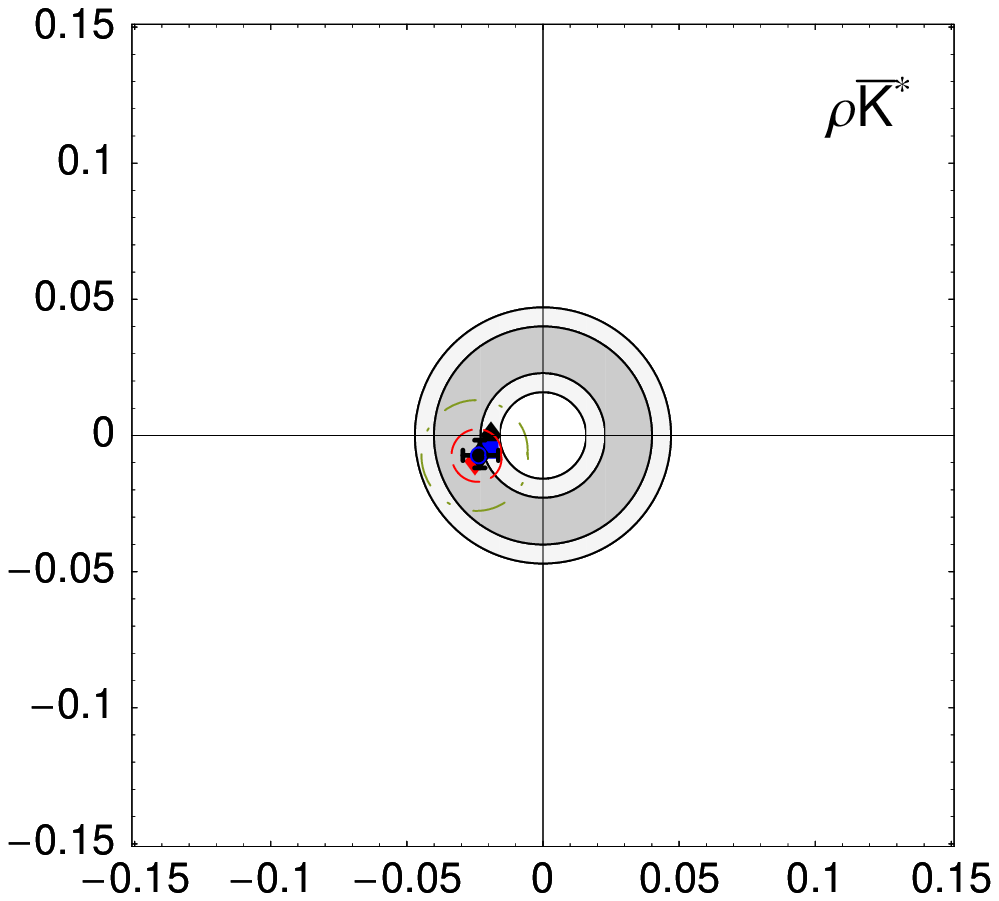}
}
\caption{\label{fig:PTpik} 
Comparing the $PP$, $PV$, $VP$, and $VV$ penguin amplitudes to data.
The figures in the upper row show theory predictions for
$\hat \alpha_4^c(M_1 M_2)/(\alpha_1(\pi \pi) + 
\alpha_2(\pi \pi))$; those in the lower row show
$\hat \alpha_4^c(M_1 M_2)/(\alpha_1(\rho \rho) + \alpha_2(\rho \rho))$.
Where available, the ranges for modulus and phase extracted from
data are also depicted.
See text for explanations. }
\end{figure}
%

We recall that in physical decay amplitudes the
parameters $a_4^p$, $a_6^p$, and
the penguin annihilation amplitude $\beta_3^p$ always appear in the
same linear combination~\cite{Beneke:2003zv}
\be
    \hat \alpha_4^p(M_1 M_2) = a_4(M_1 M_2) \pm r_\chi^{M_2} a_6(M_1 M_2)
    + \beta_3^p(M_1 M_2) ,
\ee
where the upper (lower) sign corresponds to the case $M_1=P\,(V)$,
and cannot individually be confronted with experiment. The
annihilation contribution $\beta_3^p$ is incalculable in 
factorization, since it contains an endpoint divergence; we therefore 
resort to the model defined in~\cite{Beneke:2001ev}, 
parameterizing it in terms
of a complex parameter
\be
    X_A = \ln \frac{m_B}{\Lambda_h} \, (1 + \varrho_A e^{i \phi_A})
\ee
with $\Lambda_h = 500\, \mbox{MeV}$ a hadronic scale and 
$\varrho_A=0$ by default. An additional
theoretical error is assigned to any observable by setting 
$\varrho_A=1$ and allowing the phase $\phi_A$ to take arbitrary values. 
It turns out that this is almost always by far the largest theoretical 
uncertainty for direct CP asymmetries, and among the largest 
uncertainties for branching fractions of penguin-dominated 
decays, especially those with vector mesons in the final state.  
The expressions for the annihilation amplitudes are given 
in~\cite{Beneke:2003zv} and, for $VV$ final states, 
in~\cite{Kagan:2004uw}.

To assess the validity of the QCD factorization framework, 
we show in Figure~\ref{fig:PTpik} the ratios 
$\hat \alpha_4^c(M_1 M_2)/(\alpha_1(\pi \pi) +
\alpha_2(\pi\pi))$ ($M_1M_2=\pi \bar K,\, \pi \bar K^*$)
and $\hat \alpha_4^c(M_1 M_2)/(\alpha_1(\rho \rho) +
\alpha_2(\rho\rho))$ ($M_1M_2=\rho \bar K,\,
\rho \bar K^*$),
the first two of which have been previously considered in~\cite{Beneke:2003zv}.
(For $\rho \bar K^*$ and $\rho\rho$, 
only the longitudinal polarization amplitude
is considered in the following.)
The result of the calculation is represented by the dark point 
with error bars. The nearly circular contours around this point 
show the variation of the theoretical prediction when 
the phase of the annihilation model is varied from 0 to $2\pi$ 
for fixed $\varrho_A=1,2,3$ (inner to outer circles). The blue 
square corresponds to the parameter set G, which is defined 
by $\lambda_B=200\,\mbox{MeV}$, $a_2^\pi=0.3$, $m_s=80\,\mbox{MeV}$, 
$\varrho_A=1$ and $\phi_A=-55^\circ\, (PP)$, 
$\phi_A=-20^\circ\, (PV)$, $\phi_A=-70^\circ\, (VP)$, $\varrho_A=0$
($VV$), following the favoured parameter set S4 of  \cite{Beneke:2003zv}.

As far as data is available, the amplitude ratios can be obtained
with little theory input from the well measured (CP-averaged, longitudinal)
branching fractions $\mbox{Br}(B^- \to \pi^- \bar K^{(*)0})$,
$f_L(\rho^- \bar K^{*0}) \cdot \mbox{Br}(B^- \to \rho^- \bar K^{*0})$,
$\mbox{Br}(B^- \to \pi^- \pi^0)$, 
$f_L(\rho^- \rho^0) \cdot \mbox{Br}(B^- \to \rho^- \rho^0)$,
and the rate and direct CP asymmetry
in $\bar B^0 \to \pi^+ K^{(*)-}$ (cf.~\cite{Beneke:2003zv}, (75) and
(77)). The darker rings are due to the experimental errors in
the branching fractions and the lighter ones include also the
uncertainty of $|V_{ub}|$. The angles of the wedges involve the
CP asymmetry measurements (darker region) while the lighter region
also includes the error on $\gamma$.\footnote{In calculating the 
wedges, an additional, small theory uncertainty on the ratio
$\alpha_1(\pi \bar K^{(*)0})/(\alpha_1(\pi \pi) + \alpha_2(\pi \pi))$
is not included.}
The wedges opening to the right
are ruled out (or at least disfavoured) by the fact that the measured
values of the Fleischer-Mannel ratios~\cite{Fleischer:1997um}
\be
      R_{\rm FM}^{\pi K} = \frac{\Gamma(\bar B^0\to \pi^+ K^-)}
                    {\Gamma(B^-\to\pi^- \bar K^0)}
                    = 0.91 \pm 0.07,
\qquad
      R_{\rm FM}^{\pi K^*} = \frac{\Gamma(\bar B^0\to \pi^+ K^{*-})}
                    {\Gamma(B^-\to\pi^- \bar K^{*0})}
                    = 0.93 \pm 0.19 
\ee
are both less than unity.

Three messages can be read off from the figure. (1) The magnitudes and 
phases predicted for the amplitude ratios agree reasonably 
well with data, as indicated by the error bars and the small onion-shaped 
regions. A large annihilation amplitude is disfavoured, 
since it would require fine-tuning of the phase to satisfy the 
experimental constraints, but some annihilation contribution appears to 
be required, especially for the $PV$ amplitude. (The
apparent smallness of the annihilation uncertainty in the $\rho \bar K^*$ plot
is due to cancellations in the crude annihilation model employed
here.)
There is a tendency of the predicted magnitude of the penguin
amplitude
without annihilation to be smaller than the data. This difference
is about $0.02$ to $0.03$ independent of the spins of the final-state
mesons, except for the $VV$ final state where no such difference exists.
(2) The magnitude of the penguin amplitude 
is predicted much smaller in the case containing a vector meson in the 
final state, either due to the smallness of $a^p_6$ ($\pi K^*,\, \rho K^*$) or 
a cancellation of $a^p_4$ and $a^p_6$ ($\rho K$). This is also 
reflected by the data. (3)  A priori the ratio could have lain anywhere
in the complex plane shown in the figure. The agreement found hence
constitutes a highly non-trivial check of the qualitative and
quantitative predictions for penguin amplitudes in the factorization 
framework. In this graphical representation the well-known difficulty 
to account for the small direct CP asymmetry in $\bar B^0 \to \pi^+ K^{-}$ 
in QCD factorization 
is seen as a small offset of the theoretical calculation from 
the left wedge in the first panel of the figure.

\subsection{$P/T$ and $C/T$}
\label{sec:PTCT}

Penguin-to-tree and other ratios are 
also of phenomenological importance in
$B\to \pi \pi$, $B\to \pi \rho$, $B\to \rho \rho$, and other decays.
The ratio $P_{\pi\pi}/T_{\pi\pi}$, for instance, can be determined
solely from the time-dependent CP asymmetry in $B \to \pi^+
\pi^-$ for given values of the well measured mixing phase $\phi_d$, equal to
$2\beta$ in the Standard Model, and the CKM angle $\gamma$. Conversely
theoretical predictions for these ratios allow to extract the angle
$\gamma$. Similar relations hold for final states $\pi^\mp \rho^\pm$,
$\rho^+ \rho^-$, $\pi^+ K^{(*)-}$, etc. Here we present numerical
values for a number of these ratios.

\begin{table}[t]
\renewcommand{\arraystretch}{1.4}
\begin{center}
\begin{tabular}{cccc}
Ratio & Value/Range & Value G \\
\hline
&&\\[-0.5cm]
$\displaystyle \frac{P_{\pi\pi}}{T_{\pi\pi}}$ & 
$-0.122^{+0.033}_{-0.063} + (-0.024^{+0.047}_{-0.048})i$ &
$-0.162+0.022 i$\\[0.3cm]
$\displaystyle \frac{P_{\rho\rho}}{T_{\rho\rho}}$ & 
$-0.036^{+0.006}_{-0.009} + (-0.009^{+0.007}_{-0.007})i$&
$-0.037-0.009 i$\\[0.3cm]
$\displaystyle \frac{P_{\pi\rho}}{T_{\pi\rho}}$ & 
$-0.037^{+0.015}_{-0.028} + (-0.005^{+0.024}_{-0.024})i $ &
$-0.070+0.006 i$\\[0.3cm]
$\displaystyle \frac{P_{\rho\pi}}{T_{\rho\pi}}$ & 
$0.042^{+0.039}_{-0.023} + (0.004^{+0.030}_{-0.030})i$ &
$\phantom{-}0.051-0.024 i$\\[0.3cm]\hline
&&\\[-0.5cm]
$\displaystyle \frac{C_{\pi\pi}}{T_{\pi\pi}}$ & 
$0.363^{+0.277}_{-0.156} + (0.029^{+0.166}_{-0.103})i $ &
$\phantom{-}0.691+0.165 i $ \\[0.3cm]
$\displaystyle \frac{C_{\rho\rho}}{T_{\rho\rho}}$ & 
$\phantom{-}0.198^{+0.233}_{-0.150} + (-0.009^{+0.145}_{-0.097})i$ &
$\phantom{-}0.344+0.042 i$ \\[0.3cm]
$\displaystyle \frac{C_{\pi\rho}}{T_{\pi\rho}}$ & 
$\phantom{-}0.250^{+0.229}_{-0.143} + (-0.012^{+0.127}_{-0.090})i$ &
$\phantom{-} 0.467+0.071 i$ \\[0.3cm]
$\displaystyle \frac{C_{\rho\pi}}{T_{\rho\pi}}$ & 
$\phantom{-}0.134^{+0.199}_{-0.156} +  (-0.024^{+0.152}_{-0.117})i$ &
$\phantom{-}0.283+0.138 i$ \\[0.3cm]\hline
&&\\[-0.5cm]
$\displaystyle \frac{T_{\rho\pi}}{T_{\pi\rho}}$ & 
$0.869^{+0.275}_{-0.207} + (0.014^{+0.058}_{-0.057}) i$ &
$\phantom{-}0.945-0.004 i$
\end{tabular}
\end{center}
\caption{\label{tab:ampratios}
Amplitude ratios for the $\pi\pi$, $\rho\rho$ and $\pi\rho$ final 
states. In the case of $\rho\rho$ the ratios of longitudinal polarization 
amplitudes are given. The third column gives the preferred-parameter-set G 
value.}
\end{table}

In such phenomenological studies it is convenient to define the 
colour-allowed tree amplitude $T$, the colour-suppressed 
tree amplitude $C$, and the penguin amplitude $P$ as the hadronic 
amplitudes multiplying the different CKM structures in the 
decay amplitude. In this convention, the name of an amplitude 
derives from its leading contribution, such that 
$T\sim \alpha_1$, $C\sim \alpha_2$, and $P\sim \alpha_4^c$, 
but sub-leading terms can make an important difference. For the 
following discussion of $B\to \pi\pi, \rho\rho, \pi\rho$ decays, 
we define $T,C,P$ through
\begin{eqnarray}
{\cal A}_{\overline{B}^0\to \pi^+\rho^-} 
&\propto& V^*_{ud} V_{ub} T_{\pi\rho} +  V^*_{cd} V_{cb} P_{\pi\rho},
\nonumber\\
{\cal A}_{\overline{B}^0\to \pi^-\rho^+} 
&\propto& V^*_{ud} V_{ub} T_{\rho\pi} +  V^*_{cd} V_{cb} P_{\rho\pi},
\nonumber\\
-2 {\cal A}_{\overline{B}^0\to \pi^0\rho^0} 
&\propto& V^*_{ud} V_{ub} [C_{\pi\rho}+C_{\rho\pi}] +  
V^*_{cd} V_{cb} [\ldots].
\end{eqnarray}
These definitions are related to the $\alpha_i^p$ (and annihilation) 
amplitudes by comparing the above equations to \cite{Beneke:2003zv}, 
(A.13). The proportionality factor is the same in all three 
lines and therefore irrelevant, since we shall only consider ratios. 
Analogous definitions apply to $\pi\pi$ and $\rho\rho$. 

Our results for the amplitude ratios are given in 
Table~\ref{tab:ampratios}. The first column of numbers provides 
the default result with errors, the second column the value 
in parameter set G that we currently consider as our ``best'' 
result. The theoretical calculation has significant uncertainties.
The dominant error of $P/T$ always arises from the weak annihilation model
and cannot be reduced by further calculations. The error due to
neglected higher-order corrections is difficult to estimate, but there
may be a sizable shift of the central values of $P/T$ due to the
uncalculated ${\cal O}(\alpha_s^2)$ correction to the power-suppressed
``scalar penguin'' amplitude $a_6^p$. That correction, like $a_4^p$
discussed above, involves the large Wilson coefficient $C_1$ at one loop, but
there may not be a numerical cancellation analogous to the one
observed for $a_4^p$ in Section~\ref{sec:alpha4p}.
For $C/T$, insufficiently known input parameters such as 
$\lambda_B$ and $a_2^M$, and to a lesser extent the twist-3 
correction $X_H$, are responsible for the bulk of the 
uncertainty.

There is a clear hierarchy of the penguin-to-tree ratios 
that has already been discussed in~\cite{Beneke:2003zv} and 
in the previous subsection. The difference in $C/T$ for 
the various final states is in fact a reflection of the 
same hierarchy of penguin amplitudes. While $a_{1,2}$ are 
roughly the same for $\pi\pi$, $\pi\rho$, $\rho\rho$ 
(longitudinal polarization), the amplitudes $T,C$ contain 
the up-penguin amplitude, such that $T=\alpha_1+\alpha_4^u+\ldots$, \,
$C=\alpha_2-\alpha_4^u+\ldots$. It is worth noting that 
$C/T$ can acquire a significant imaginary part
from the 1-loop vertex and spectator-scattering correction, 
and the large phase of the up-penguin amplitude. 
It will be interesting to detect these characteristic 
features in the experimental data. 

\subsection{Relating electroweak penguin to tree amplitudes}
\label{sec:EWP}
The electroweak penguin amplitudes $\alpha^p_{3,\rm EW}$, 
$\alpha^p_{4,\rm EW}$ are related to the tree amplitudes 
$\alpha_{1,2}$ under certain assumptions 
\cite{Neubert:1998pt,Gronau:1998fn,Buras:1998rb}. Therefore, more than 
the calculated values of the electroweak amplitudes, the deviations  
from these relations are of interest. The assumptions made in 
deriving the relations are the neglect of the electroweak penguin 
operators $Q_{7,8}$, since they have small Wilson coefficient 
compared to $Q_{9,10}$ at the scales of interest, and SU(3) 
flavour symmetry, when a relation between $\pi\pi$ and $\pi K$ 
final states is involved. In addition there are further assumptions, 
which amount to neglecting the charm and bottom content of the operators 
$Q_{9,10}$, and to not considering penguin contractions. Since 
our calculation does not make use of any of these assumptions, 
it is interesting to investigate how well these widely used 
amplitude relations are satisfied. We confine ourselves to the 
discussion of the $\pi K$ and $\pi\pi$ final states.

The most solid relation is \cite{Neubert:1998pt}
\be
\delta_\mathrm{EW} = -\frac{3}{2}
   \left|\frac{V^*_{cs}V_{cb}}{V^*_{us} V_{ub}}\right| \frac{R_{\pi K} \alpha^c_{3, \rm EW}(\bar K\pi)+
\alpha^c_{4,\rm EW}(\pi \bar K)}{\alpha_1(\pi\pi)+\alpha_2(\pi\pi)} 
\approx -\frac{3}{2}\left|\frac{V^*_{cs}V_{cb}}{V^*_{us} V_{ub}}\right|
\frac{C_9+C_{10}}{C_1+C_2}
\label{ewp1}
\ee
with $R_{\pi K}=f_\pi F^{BK}(0)/(f_K F^{B\pi}(0))$ and a similar 
relation with $\pi\pi$ in the numerator and $R_{\pi K}\to 1$. Let 
${\cal R}_{\rm NR}$ be the ratio of the middle expression
to the right-hand side. We find 
\be 
{\cal R}_{\rm NR} = (1.02^{+0.27}_{-0.22})\,e^{i(1\pm 1)^\circ}
\ee
The closeness to 1 involves a cancellation between an SU(3) breaking
on the order of $10 \%$ and an error of similar size due
to neglecting the electroweak penguin operators  $Q_{7,8}$
(for $\pi\pi$ we would find
${\cal R}_{\rm NR}\approx (0.92^{+0.05}_{-0.04})\,e^{i(1 \pm 1)^\circ}$).
The effect of penguin contractions is below the 
$(1-2)\%$ level. Note also that the ratio remains real to an 
excellent approximation. The relation holds to much better accuracy
than at the time when~\cite{Beneke:2001ev} was written, 
where it was studied without the 1-loop spectator 
scattering correction. The difference with respect to~\cite{Beneke:2001ev} 
is almost exclusively due to
the change in the hadronic parameter ratio $R_{\pi K}$, which is also
responsible for the bulk of the error.

A second relation follows from SU(3) symmetry \cite{Gronau:1998fn}, 
which also involves annihilation amplitudes. Neglecting such 
power-suppressed amplitudes, we obtain
\be
\frac{R_{\pi K} \alpha^c_{3,\rm EW}(\bar K\pi)-
\alpha^c_{4,\rm EW}(\pi \bar K)}{\alpha_1(\pi\pi)-\alpha_2(\pi\pi)} 
= \frac{C_9-C_{10}}{C_1-C_2}
\label{ewp2}
\ee
Let ${\cal R}_{\rm GPY}$ be the ratio of the left-hand to the 
right-hand side. We find 
\be 
{\cal R}_{\rm GPY} = (1.29^{+0.58}_{-0.38})\,e^{i(- 1^{+ 3}_{-2})^\circ}
\ee
The deviation from 1, which is bigger than in ${\cal R}_{\rm NR}$,
comes about because now the SU(3)-breaking correction
(for $\pi\pi$ we would find  ${\cal R}_{\rm GPY}\approx 1.13$) and the effect of 
neglecting $C_{7,8}$ add up. Individually, both effects are nearly 
twice as large as in the first relation, which is also reflected 
in the larger error. Nevertheless, ${\cal R}_{\rm GPY}$ 
remains real to an excellent approximation.

The previous two relations (\ref{ewp1}), (\ref{ewp2}) hold 
under the stated assumptions irrespective of the values 
of the Wilson coefficients 
$C_{9,10}$. Observing that 
\be
\frac{C_9+C_{10}}{C_1+C_2} \approx \frac{C_9-C_{10}}{C_1-C_2} 
\approx -0.00896 \equiv \kappa,
\ee
at $\mu_b=4.8\,\mbox{GeV}$ and neglecting the $1\%$ difference between 
the first three items in this expression, 
stronger relations are obtained in \cite{Gronau:1998fn},
\be
\alpha^c_{3,\rm EW}  = \kappa \alpha_1,
\qquad \alpha^c_{4,\rm EW} = \kappa \alpha_2,
\ee
where due to the SU(3)-symmetry assumption the arguments $M_1M_2$ of 
the $\alpha_i$ can be any out of $\pi\pi$, $\pi \bar K$, $\bar K\pi$ and 
$K \bar K$, $\bar K K$. The first of these relations is 
extraordinarily well respected,
\be
\frac{\alpha^c_{3,\rm EW}(\pi\pi)}{\kappa \alpha_1(\pi\pi)} 
= (1.00 \pm 0.01)\,e^{i(0^{+0}_{- 1})^\circ}, 
\ee
because it involves the colour-allowed amplitudes, and 
the effect of $Q_{7,8}$ is very small. This holds for any of the 
above final states $\pi\pi, \pi K, KK$.
In contrast, the 
second relation between the colour-suppressed amplitudes 
is poor. We find  
\be
\frac{\alpha^c_{4,\rm EW}(\pi\pi)}{\kappa \alpha_2(\pi\pi)} 
= (0.64^{+0.25}_{-0.24})\,e^{i(3^{+11}_{-14})^\circ}.
\ee
Our theoretical expectations should be useful to estimate the error 
incurred by employing these relations in data-driven approaches 
to non-leptonic decays. 

\section{Conclusion}
\label{sec:conclude}

This paper, together with \cite{Beneke:2005vv}, completes the 
calculation of 1-loop 
spectator-scattering corrections to all flavour-nonsinglet, leading-power 
decay amplitudes in non-leptonic $B$ decays. 
The calculation shows that factorization works technically: 
the infrared singularities cancel in the matching calculation, 
and the convolution integrals converge at their endpoints. 

The 1-loop corrections are numerically well-behaved. As a general 
rule, we find significant effects for the colour-suppressed amplitudes 
$\alpha_2$, $\alpha_3^p$ and $\alpha^p_{4,\rm EW}$, and small 
effects for the others. In case of the QCD penguin amplitude 
$a_4^p$ this conclusion is reached only as the result of a 
numerical cancellation in 
the terms proportional to the large Wilson coefficient $C_1$. 
As a consequence, there is little impact of the newly calculated
spectator-scattering corrections to penguin amplitudes on the
phenomenology of non-leptonic branching fractions and CP asymmetries.
Further improvement 
of the calculation now requires the calculation of the 2-loop 
vertex corrections to the form-factor term in the factorization 
formula~\cite{Bell06}, and an understanding of the power-suppressed 
but large penguin amplitude $a_6^p$. 

In the final section of the paper we investigated a few amplitude 
ratios that play an important role in determining $\gamma$ ($\alpha$) 
from tree-dominated $b\to d$ transitions. Vice versa, assuming 
a value of $\gamma$, these ratios may be determined from data 
and compared to the theoretical calculation. 
A detailed discussion of branching fractions and CP asymmetries 
with account of 1-loop spectator scattering will be presented 
elsewhere. 

\subsubsection*{Acknowledgements}

This work is supported in part by the 
DFG Sonder\-forschungs\-bereich/Trans\-regio~9 
``Com\-puter\-ge\-st\"utz\-te Theoretische Teilchenphysik''.

\end{document}